\documentclass[manuscript]{aastex}

\usepackage{graphicx}
\usepackage{dcolumn}
\usepackage{paralist}
\usepackage{comment}
\usepackage{graphicx,epsfig}
\usepackage{multirow}
\usepackage{wrapfig}
\usepackage{float}

\slugcomment{To be sumitted to PASP.}

\shorttitle{NEO Observation with Synthetic Tracking}
\shortauthors{C.Zhai et al.}

\def\beq{\begin{equation}}
\def\eeq{\end{equation}}
\def\beqa{\begin{eqnarray}}
\def\eeqa{\end{eqnarray}}
\def\as{^{\prime\prime}}
\def\am{^\prime}
\begin{document}
\title{Near-Earth Object Observations using Synthetic Tracking}

\author{Chengxing Zhai\altaffilmark{1,5}, Michael Shao\altaffilmark{1}, Navtej Saini\altaffilmark{1}, Philip Choi\altaffilmark{2}, Nez Evans\altaffilmark{2}, Russell Trahan\altaffilmark{1}, Kutay Nazli\altaffilmark{3,4},  Max Zhan\altaffilmark{1}}
\email{cxzhai@hkust.hk}
\altaffiltext{1}{Jet Propulsion Laboratory, California Institute of Technology, 4800 Oak Grove Dr, Pasadena, CA 91109}
\altaffiltext{2}{Department of Physics, Pomona College, Claremont, CA, 91711}
\altaffiltext{3}{Mathematical Institute, Leiden University, Snellius Gebouw, Niels Bohrweg 1, NL-2333 CA Leiden, The Netherlands}
\altaffiltext{4}{Leiden Observatory, Leiden University, Oort Gebouw, Niels Bohrweg 2, NL-2333 CA Leiden, The Netherlands}
\altaffiltext{5}{Division of Emerging Interdisciplinary Areas, The Hong Kong University of Science and Technology, Clear Water Bay, Hong Kong}
\begin{abstract}
Synthetic tracking (ST) has emerged as a potent technique for observing fast-moving near-Earth objects (NEOs), offering enhanced detection sensitivity and astrometric accuracy by avoiding trailing loss. This approach also empowers small telescopes to use prolonged integration times to achieve high sensitivity for NEO surveys and follow-up observations. In this study, we present the outcomes of ST observations conducted with Pomona College's 1 m telescope at the Table Mountain Facility and JPL's robotic telescopes at the Sierra Remote Observatory. The results showcase astrometric accuracy statistics comparable to stellar astrometry, irrespective of an object's rate of motion, and the capability to detect faint asteroids beyond 20.5th magnitude using 11-inch telescopes.

Furthermore, we detail the technical aspects of data processing, including the correction of differential chromatic refraction in the atmosphere and accurate timing for image stacking, which contribute to achieving precise astrometry. We also provide compelling examples that showcase the robustness of ST even when asteroids closely approach stars or bright satellites cause disturbances. Moreover, we illustrate the proficiency of ST in recovering NEO candidates with highly uncertain ephemerides.

As a glimpse of the potential of NEO surveys utilizing small robotic telescopes with ST, we present significant statistics from our NEO survey conducted for testing purposes. These findings underscore the promise and effectiveness of ST as a powerful tool for observing fast-moving NEOs, offering valuable insights into their trajectories and characteristics. Overall, the adoption of ST stands to revolutionize fast-moving NEO observations for planetary defense and studying these celestial bodies.
\end{abstract}

\section{Introduction}
Observing near-Earth Objects (NEOs) holds significant importance for planetary defense, solar system formation studies, and resource mining applications. 
While meter-size or smaller NEOs harmlessly disintegrate in the Earth's atmosphere, larger ones can cause devastating damage. 
The US Congress mandated NASA to find NEOs larger than 140 m with at least 90\% completeness due to the potential regional devastation caused by such impacts \citep{NSTC2023}. 
NASA's Near-Earth Object Observations programs (NEOO) fund projects to discover, track, and characterize NEOs in response to this mandate.
Currently, we have identified approximately 40\% of NEOs larger than 140 m, leaving about 15,000 NEOs to be discovered \citep{NSTC2023}.
Current surveys, such as the Catalina Sky Survey (CSS) and PanSTARRS, are producing more than 3000 NEOs per year (https://cneos.jpl.nasa.gov/stats/site\_all.html) with about 500 larger than 140 m. While the detection rate has been
steadily increasing, finding 90\% of the NEOs of size 140 m or larger can easily take an extra 20 years.
Fortunately, the upcoming Rubin Telescope and NEO Surveyor Mission are expected to accelerate the discovery process \citep{NSTC2023}.

However, we cannot be optimistic because
NEOs smaller than 140 m can still be very hazardous and the frequency for smaller asteroids to impact Earth is much higher than that of larger asteroids \citep{NSTC2023}.
The incident of the Chelyabinsk meteor \citep {Brumfiel2013} measuring about 20 m underscores the need to detect potential threats from NEOs larger than 10 meters. Therefore, 
NEOO seeks to inventory all the NEOs that could post a threat or serve as potential mission targets.
NEOs smaller than 140 m constitute a much larger population \citep{Tricarico2017} with the vast majority of their threats remaining unknown \citep{NSTC2023} because their smaller sizes
require closer proximity to Earth to be sufficiently bright for observation. The associated trailing loss from the faster motion rate becomes a substantial hurdle for surveying small hazardous NEOs.

Synthetic tracking (ST) is a powerful technique designed to detect fast-moving NEOs and perform
follow-up observations \citep{Shao2014, Zhai2014,Heinze2015}.
Enabled by CMOS cameras and modern GPUs, ST takes multiple short-exposure images to avoid trailing loss associated with traditional long exposure ($\sim$ 30-second) CCD images
and integrates these short-exposure images in post-processing using GPUs.
CMOS cameras can read large format frames ($\sim$60 Mpixel) at high frame rates with read noise of only about 1e per read%
\footnote{See \url{https://www.qhyccd.com/astronomical-camera-qhy600/}, \url{https://www.ximea.com/en/products/xilab-application-specific-custom-oem/scientific-scmos-cameras-with-front-back-illumination} and \url{https://www.photometrics.com/products/scmos/} for more information.}.
Such a low read noise means even during dark times near the new moon, taking frames at 1 Hz, the read noise is still lower than the sky background noise for an 11-inch telescope.
ST avoids trailing loss using a high frame rate (short exposure time) to make NEO motion negligible compared with the size of the point-spread-function (PSF), which is typically 2 arcsec. 
For surveying NEO, a 1 Hz frame rate is usually sufficient to avoid trailing loss because NEOs would unlikely move more than a typical PSF size of 2 arcsec during 1 sec%
\footnote{A higher rate would be needed for observing a NEO gets very close to the Earth or satellites.}.

A single short exposure image in general does not suffice for detecting new NEOs, therefore,
we need to integrate many frames (of order 100) to improve the signal-to-noise ratio (S/N)
in post-processing. For follow-up observations, this task can be readily carried out because we know
approximately the rate of motion, so in post-processing, we can stack up the images according to the motion to track the target. Even though the rate of motion may not be very accurate in case the ephemeris is off,
the effort to find the best tracking can be made by adjusting the tracking with a least-squares fitting.
For detecting new NEOs, this effort of post-processing integration is very large because we need to search over a large set of trial velocities, which is typically a 100$\times$100 velocity grid for us.
To speed up this, ST uses modern GPUs that offer thousands of processors at a low cost. For example,
using the Nvidia V100 GPUs, we can keep up with real-time processing
for our NEO survey experiments using an exposure time of 5 seconds.
ST has demonstrated success in detecting small NEOs of $\sim$ 10 m. These objects tend to move fast ($>$ 0.5 arcsec/sec) and often elude
surveys like PanSTARRS and CCS due to the excessive trailing loss.
With the capability of integrating a long time (many frames), ST empowers small telescopes to detect faint objects, a feat unattainable without this technique.

The flexibility of ST post-processing has many advantages over the traditional long-exposure approach.
ST can track both the target and stars, thus, producing more accurate astrometry than the traditional approach that has to deal with centroiding streaked objects leading to degraded precision as the rate increases \citep{Veres2012}. 
We have demonstrated 10 mas level NEO accuracy using ST \citep{Zhai2018} with typically better than 10 mas astrometric solutions. To achieve 10 mas level NEO astrometry, we found it necessary to correct the differential chromatic refraction (DCR)
effect of the atmosphere to account for the wavelength dependency of air refraction.

In addition, we found ST robust against star confusion in performing follow-up observations, where we can exclude the frames where the NEO gets very close to a star.
The chance of confusion increases with the rate of motion, so traditionally it would be hard to avoid the contamination of the streaked stars when tracking fast-moving NEOs without using ST.
Another advantage of using ST is its proficiency in recovering NEOs with highly uncertain ephemerides, where the significant rate errors would make the traditional approach fail to track these NEOs.

This paper presents the results and data processing from using ST for NEO observation. The paper is organized as follows: In Section 2,
we describe the instrument, operations, and data processing involved in using ST to observe NEOs. In Section 3, we present results showcasing the advantages of employing ST.
Finally, we conclude with an outlook on the future of ST in NEO observation.

\section{NEO Observation using Synthetic Tracking}
ST necessitates the capture of images with an exposure time short enough to prevent significant NEO motion relative to the size of the PSF. However, this must be balanced with the potential increase of read noise from reading out images too rapidly. Consequently, determining the ideal exposure time and the number of frames becomes a critical decision, influenced by the system's hardware configuration and the prevailing sky background level. In this section, we offer a comprehensive overview of our instrumentation and elaborate on the operational strategies, with a specific emphasis on the meticulous design of observation cadence.

 \subsection{Instrument Description}
Our NEO observations use a total of three systems, each with distinct key parameters outlined in Table~\ref{table:1}.
\begin{table}
\caption{System Parameters\label{table:1}}
\centering
\begin{tabular}{|c|c|c|c|}
\hline
\textbf{Location (Obs. code)} & \textbf{TMF (654)} & \textbf{SRO 1(U68)} & \textbf{SRO 2 (U74)} \\
\hline
Primary Diameter (inch) & 40 & 11 & 14 \\
\hline
Number of Telescopes & 1 & 3 & 1 \\
\hline
focal length(m) & 9.6 & 0.62 & 0.79\\
\hline
Detector & Photometrics 95B-25mm & QHY/ZWO 60M & ZWO 60M\\
\hline
QE, Peak/Average & 0.95/0.8 & 0.95/0.8 & 0.95/0.8\\
\hline
Pixel size (um) & 11 & 3.76 & 3.76 \\
\hline
Pixel Scale (as) & 0.226 & 1.26 & 0.98 \\
\hline
Read noise (e) & 1.6 & 1-2 & 1-2 \\
\hline
Dark current (e/sec) & $<$ 1 (T = $0^\circ$ C) & $< $0.5 (T = $0^\circ $C) & $< $0.5 (T = 0$^\circ$ C) \\
\hline
Highest Frame Rate (fps) & 30 & 2.5 &2 \\
\hline
Array size & 1608$\times$1608 & 9576$\times$6388 & 9576$\times$6388 \\
\hline
Field of View (deg$\times$deg) & 0.1$\times$0.1 & 2.2$\times$3.3 & 1.7$\times$2.6\\
\hline
Typical seeing (as) & 2 & 2 & 2\\
\hline
Sky darkness (mag/as$^2$) & 20 & 21 & 21\\
\hline
\end{tabular}
\end{table}
The first system comprises Pomona Coellge's 40 inch telescope located at the Table Mountain Facility (TMF). This Cassegrain telescope features a 1 m f/2 primary mirror with a 30 cm secondary mirror, resulting an effective focal length of 9.6 meter for the imaging system.
A Photometrics 95B Prime sCMOS detector is installed at the Cassegrain focus with a pixel array size of 1608$\times$1608. The 11$\mu$m pixel corresponds to a scale of 0.226$\as$ per pixel enabling a critical sampling of PSF for our best seeing conditions of 1.5 as at the TMF.  The field of view (FOV) is $6\am{\times}6\am$. 
This system does not have any refractive elements, thus its field distortion
 is insensitive to color making astrometric calibration easier. We have used it to achieve 10 mas level NEO astrometry \citep{Zhai2018}.

We have built two additional robotic telescope systems using commercial off-the-shelf (COTS) telescopes from Celestron located at the Sierra Remote Observatory (SRO).
One system (SRO1) consists of three 11-inch RASA telescopes at f/2.2 arranged with offsets in Declination, giving approximately a total FOV of 6.6 deg $\times$ 3.3 deg. We use SRO1 to survey NEOs nominally.
The other system (SRO2) has a single 14-inch RASA telescope for follow-up observations. We use both the ZWO and QHY 600 Mpixel CMOS cameras using the Sony IMX 455 Chip, which has a pixel size of 3.76$\mu$m giving 
pixel scales of 1.26as and 0.98as respectively for the SRO1 11-inch and SRO2 14-inch telescopes. The relevant parameters are listed in Table~\ref{table:1}.
 
\subsection{Operation}
\subsubsection{Configuring Science Observations\label{sec:sci_obs}}
For operation, we want to maximize the instrument S/N in determining the exposure time and number of frames to acquire.
To minimize trailing loss, the exposure time should be as short as possible. However, for the same amount of integration time,
using shorter exposure increases the number of reads, thus the read noise.
We now discuss how to choose an appropriate exposure time to balance the trailing loss and the total amount of noise.

The total background noise per pixel can be modeled as
the RSS (root-sum-squares) of the read noise, dark current, and background illumination:
\beq
  \sigma_n = \sqrt{\sigma_{\rm rn}^2 + \Delta t( I_{\rm bg} + I_{\rm dark})} \,,
\eeq
where $\sigma_{\rm rn}$ is the standard deviation of read noise, $\Delta t$ is the exposure time, $I_{\rm bg}$ is the sky background, and $I_{\rm dark}$ is the detector dark current.
It is convenient to define a time scale $\tau_2$ for the variance of the read noise to be the same as that of the noise from the background illumination plus the dark current as
\beq
   \tau_2 \equiv \sigma_{\rm rn}^2 /(I_{\rm dark} + I_{\rm bg}) \,.
\label{tau2_def}
\eeq
We can factor
\beq
\sigma_n = \sqrt{\Delta t( I_{\rm bg} + I_{\rm dark})} \sqrt{1 + {\tau_2 \over \Delta t } } \,,
\label{rn_factor}
\eeq
where the second factor shows the contribution of read noise to the total noise, which increases as we shorten the exposure time $\Delta t$. 
When $\tau_2 \ll \Delta t$, we are background noise limited, the total noise $\sigma_n$ only increases slowly when shortening exposure time $\Delta t$. When $\Delta t$ is not much larger than
$\tau_2$, the read noise factor  $\sqrt{1 + {\tau_2 / \Delta t } }$ becomes sensitive to the variation of $\Delta t$.

\begin{figure}[ht]
\epsscale{0.66}
\plotone{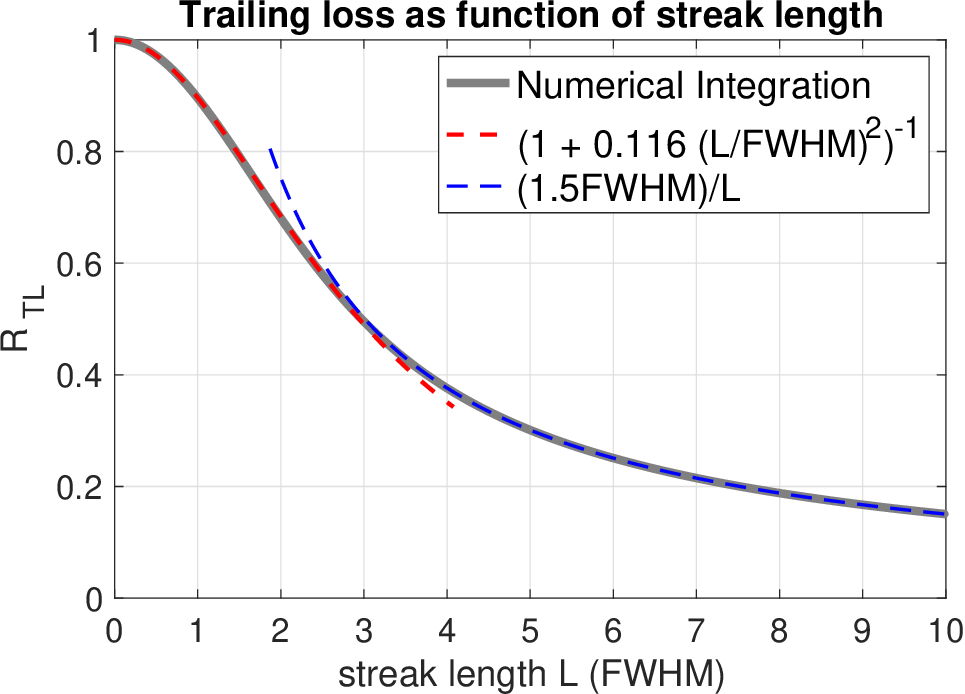}
\caption{Detection sensitivity with trailing loss as function of the streak length measured in FWHM of PSF.\label{fig:trailingLoss}}
\end{figure}

Fig.~\ref{fig:trailingLoss} illustrates the relationship between trailing loss and streak length (for detailed derivations, refer to Appendix A). The rule of thumb for using ST to observe NEOs is 
to set an appropriate exposure time $\Delta t$ so that even the most swiftly moving NEOs within the scope of interest do not result in streaks spanning more than one PSF.
This constraint effectively contains trailing loss to below 12\%.

It is useful to introduce a scale for rate of motion as $FWHM/\tau_2$, corresponding to a streak length of the PSF's FWHM for exposure time $\Delta t = \tau_2$.
If the rate range of interest is much less than $FWHM/\tau_2$, we then can easily choose an exposure time $\Delta t$  to be
larger than $\tau_2$ for pixel noise to be background noise limited and simultaneously having very little trailing loss. This is the typically the case for using a CMOS camera to observe NEOs
because typical CMOS cameras have only 1-2 e read noise when operating in rolling shutter mode (for example, see https://www.photometrics.com/products/prime-family/prime95b).
For Pomona College 40 inch telescope at TMF, the $\tau_2$ is less than $0.3 s$. Assuming PSF FWHM is 2 as, $FWHM/\tau_2 \sim 6.7 as/sec$ is much higher than typical NEO's sky rate. 

\begin{figure}[ht]
\epsscale{0.8}
\plotone{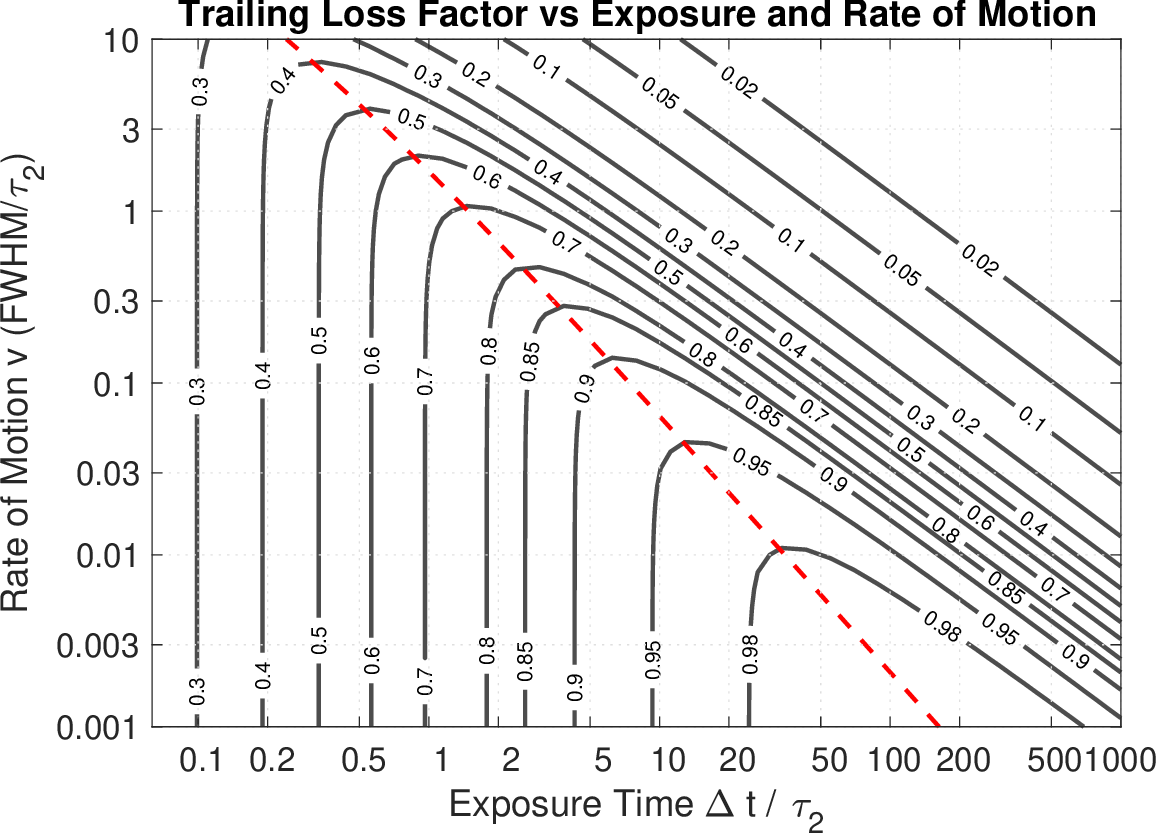}
\caption{Detection sensitivity as function of rate of motion and exposure time in contours. The red dashed lines represent optimal exposure time for a given velocity.\label{fig:sens_trade}}
\end{figure}

To optimize the sensitivity, we include the trailing loss and read noise factor together to define a detection sensitivity $S(v, \Delta t)$ as
\beq
   S(v, \Delta t) \equiv (1 + \tau/\Delta t)^{-1/2} R_{TL}(\Delta t v) \,.
\eeq
Fig.~\ref{fig:sens_trade} displays contours of constant values of $S(v, \Delta t)$ as a function of $\Delta t $ and $v$ in units of $\Delta t / \tau_2$ and FWHM/$\tau_2$ respectively.
For a given rate of motion range, there is an optimal exposure time marked by the red dashed line.
When the rate of motion $<$ FWHM/$\tau_2$, we have have a pretty good sensitivity of 0.7 with an optimal choice of exposure time $\Delta t/\tau_2 \sim 1.5$.
If rate of motion $<$ 0.3 FWHM/$\tau_2$, this can be improved to 0.85.
As an example, considering again our telescope at TMF with $\tau_2 = 0.3 s$ and FWHM $= 2$ as giving the unit for velocity FWHM / $\tau_2 \sim 2/0.3 = 6.7 (as/s)$. If we are interested in NEOs moving as fast as $1 as/sec$, or 0.15 in units of FWHM$/\tau_2$, 
we have a range of exposure times would achieve better than 0.85 detection sensitivity regarding trailing loss and read noise trade Fig.~\ref{fig:sens_trade}. This is consistent with our discussions regarding
the regime where the read noise is much lower than the background noise.

Discussion above is mainly for observing faint objects, especially for discovering new NEOs, where we have ignored the photon noises from the target itself.
In case of observing a bright target, whose photon noise is much higher than the total read noises of all the relevant pixels within a PSF, we in general only need to consider the trailing loss
to use an exposure time so that $v_{\rm max} \Delta t <$ FWHM.

After choosing an exposure time, the second factor to consider is the total integration time $T = N_f \Delta t$, or the number of frames $N_f$ to integrate for the final SNR to be sufficient for detection or achieving certain astrometric accuracy.
For example, at SRO, we require a detection threshold of $7.5$ for a low false positive rate of 2\% per camera field \citep{Zhai2014}. 
The 40 inch telescope at TMF is mainly used for follow-up observations providing highly accurate astrometry. We have been targeting at better than 100 mas accuracy, which for 
a PSF size of 2 arcsec, this means the SNR would need to be at least  13 in view of uncertainties of centroiding is $\sim 0.64$ FWHM/SNR \citep{Zhai2014}.
Using results in Appendix A, we have the total SNR 
\beq
   SNR = {\sqrt{T} I_s \over \sqrt{4 \pi \sigma_g^2} \sqrt{I_{\rm dark} + I_{\rm bg}} } S(\Delta t, v) \approx {N_{\rm target} S(\Delta t, v)  \over 1.5 \, {\rm FWHM} \sqrt{N_{\rm bg} + N_{\rm dark}}} \,,
\eeq
where have introduced total number of photoelectrons $N_{\rm bg} = T 10^{-0.4 (m_b - m_0)} a^2$ and $N_{\rm target} = T 10^{-0.4 (m_t - m_0)}$ for the sky background and target, and total dark counts $N_{\rm dark} = T I_{\rm dark}$
with $m_0$ being the telescope system zero point (stellar magnitude giving 1 photoelectron/sec), 
$a$ being the pixel scale, $m_b$ being the background brightness per pixel measured in magnitude, and $m_t$ being the target brightness.
In general, using ST, we operate with $S(\Delta t, v) > 0.7 $ for most of the NEO observations.

As an example, our 11-inch telescopes at the SRO, we use an exposure time of 5 s and a good PSF would have FWHM of 3 as, for our plate scale ($\sim$2.5 pixel).
Our 11 inch telescope's zero point is $m_0 = 22.1$. The dark current is only about 0.5 e/s and using the sky magnitude to be 20.5 mag per arcsec square.
Putting source brightness $m_t = 20.5$, $T = 500 s$, we get $\tau_2 = 1.6^2 / (0.5 {+} 1.26^2 \times 10^{-0.4 \times (20.5 - 22.1)}) \approx 0.34 s$. We are interested
in velocity range of $0.6$ as/s, or 0.07 in unites of FHWM/$\tau_2 \sim$ 9 as/s.
According to Fig.~\ref{fig:sens_trade}, the sensitivity S($\Delta t$ = 5 s, v = 0.6 as/s) $\approx$ 0.9 for using 5 second exposure.
The SNR is then
\beq
   SNR = {500 \times 10^{-0.4 \times (20.5-22.1)} \times 0.9 \over 1.5 \times 2.5 
   \sqrt{500 \times 10^{-0.4\times (20.5 - 22.1)} \times 1.26^2 + 500 \times 0.5} } \approx 8.6 \,.
\eeq

\subsubsection{Survey and Follow-up Observation Cadence}
We have been experimenting with the SRO1 system to detect new NEOs.
We use a 5-second exposure and integrate 100 frames to reach a detection limiting magnitude
of about 20.5 for dark nights near new moon (see the example of the last subsection). 
The SRO1 system has a combined FOV of about 20 sqdeg from the three RASA 11-inch telescopes.
On average, we spent about 700 seconds per pointing, which includes slew time,
refocusing time (every 8 pointings), and an extra waiting time for the synchronization of the three 
telescopes especially the extra 800-900 msec dead time that the ZWO camera has between 5-sec frames while QHY cameras do not have this dead time.
We scan along the RA four consecutive FOVs and then repeat the scan for confirmation. 
Repeating scan is operationally inefficient and we are working on a software capability 
to use the SRO2 system to do follow-up observations upon a detection from SRO1. This triggered follow-up 
allows SRO1 to scan the sky at a rate two times faster (without the burden of the revisit).

We regularly perform follow-up observations for NEO candidates from the Minor Planet Center's confirmation page (NEOCP) using the system at the TMF.
Because the telescope is sufficiently large, the frame is dominated by sky noise, i.e., $\tau_2 \sim 0.3 s$ is small relative to $\Delta t$, which
we typically use 1s, 2s, and 3s exposures and integrate. We usually integrate 300 or 600 frames depending on the brightness and rate of the target.

The 14-inch telescope system (SRO2) performs follow-up observations for candidates with large uncertainties in their ephemerides. These candidates are not suitable for the TMF 40 inch telescope to follow due to the small FOV.
For NEO candidates from SRO1, we use also 5-second exposure and a 100 frame integration.
Our SRO1 system uses S/N threshold of 7.5 to survey NEOs. The larger collecting area of 14-inch (versus 11 inch)
and better imaging quality gives us an improvement factor of about 1.6 in S/N, thus SRO2 
can reliably confirm the candidates with an S/N of 1.6$\times$7.5 = 12 unless they are false detections. 
This telescope has been also used to confirm objects from the Zwicky Transit Facility (ZTF) and
NEO candidates from the NEOCP. We are developing software to fully automate the operation of SRO2 to
schedule and perform follow-up observations, as well as processing and submitting the data.

\subsubsection{Calibration\label{sec:cal}}
For calibration, we generate a mean dark frame, a flat field response, and a list of bad pixels.
The mean dark frame is estimated by averaging multiple dark frames taken with the
same exposure time as the science data.
The flat field response, which physically is the product of the relative pixel quantum efficiency and optical throughput, 
can be measured by observing the twilight sky.
The flat field response can be computed by taking an average over multiple measurements
and then normalized so that the mean response over the whole field is 1. 
We generate a list of bad pixels by applying a noise level upper limit threshold, a
dark level upper limit threshold, and a lower limit threshold for flat field response.

The twilight flat field calibration can be performed in two ways. One setup is to take the measurements
when the twilight light is much stronger than any stars in the field so that photons from stars
in the field can be ignored relative to the sky background. To avoid saturation, we
typically use a very short exposure (no more than 0.1s) to keep the pixel light level for the twilight sky 
at about half full-well counts. Turning off the tracking to let stars drift in the field helps because
the trailing loss further reduces the star lights relative to the twilight sky background.
We usually take hundreds of frames and it is straight forward to take an average over these frames
and perform a normalization to yield a flat field response.
However, this approach requires the experiment to be carried out in a very limited time 
window during twilight.

In case of missing the desired twilight time window, 
an alternative approach for flat field response calibration can be employed. 
This approach involves activating sidereal tracking and deliberately shifting the pointing 
to capture multiple sky background images with stars at different pixel locations in the field.
This diversification guarantees that each pixel has multiple opportunities to exclusively
capture the sky background free of star-generated photons. 
Subsequently, the data is processed by first eliminating pixel data where star signals are detected.
After scaling the sky background of each sky image to the same level, an average can be
computed for each pixel across the image set, exclusively considering instances when the pixel registers the sky background without any star signals.
The data processing is slightly more involved, but we gain the flexibility of when to take the data.
This approach works even when the sky background is not high, where a longer integration can be implemented as needed.

\subsection{Data Processing}
The framework and procedural stages of data processing have been outlined in \cite{Zhai2014}. 
For follow-up observation data processing, \cite{Zhai2018} provides a thorough description of how to generate astrometry for observing known NEOs. Here we give an overview of the data processing, highlight how ST identifies targets, and detail in generating highly accurate astrometry by correcting the DCR effect of the atmosphere as well as accounting for accurate timing when stacking up frames.

\subsubsection{An overview of synthetic tracking data processing}

In the contrast to conventional asteroid detection data processing \citep{Rabinowitz1991, Stokes2000, mops2013}, ST works on a set of short-exposure images, which we call a ``datacube"
because of the extra time dimension in addition to the camera frame's row and column dimension. The goal of data processing is to 1) identify the stars in the field and
 find an astrometric solution to map sky and pixel coordinates; 2) detect significant signals (search mode) or identify target (follow-up); 3) estimate astrometry and photometry for the detected objects or follow-up target.
 
The data processing consists of three major steps:
\begin{enumerate}
\item {\it Preprocessing}, where we apply calibration data, remove cosmic ray events, and re-register frames to get data ready;
\item {\it Star field processing}, where we estimate sky background, identify stars in the field, and match stars against a catalog; 
\item {\it Target processing}, where we identify the target and estimate its location, rate of motion, and photometry.
\end{enumerate}

{\it Preprocessing} is instrument dependent and generally requires subtracting a mean dark frame of the same exposure time from each frame and then dividing each frame by a flat field response to
account for the throughput and QE variation over the field (see subsection \ref{sec:cal}). While a well-tracking system may not need re-registration, a re-registration is needed for our systems, 
which could drift more than 10 arcsec during the course of an integration. Re-registration can be done by estimating offsets between frames by estimating positions of one or a few bright stars in each frame or cross-correlating Fourier
transforms of each frame. We then remove the cosmic ray events and bad pixel signals by setting the values at these pixels to a background value.
Cosmic ray events are identified as signal spikes above random noise level localized in both temporal and spatial dimension.

{\it Star field processing} first detects stars in the field by co-adding all the frames with stars well-aligned after the frame re-registration.
A least-squares fitting to a Gaussian or Moffat PSF is used to estimate the pixel locations of reference stars. 
A planar triangle matching algorithm \citep{Padgett1997} identifies stars in the field by matching similar triangles formed by triplets of stars at the vertices from both the field and the catalog, where the shapes of triangles 
are determined by the relative distances between the stars. To reduce the computer time for matching stars, we usually
need to know in advance the approximate location of the field in the sky, the pixel scale, and the size of the FOV to look up a star catalog, which is the Gaia Data Release 2 (DR2) \citep{Gaia2016} for our data processing.
 To avoid excessive combinations of triplets of stars, this process starts with a small subset of the brightest stars in the field.
A pair of correctly matched triangles in the field and catalog gives an affine transformation between the pixel and sky coordinates. 
The affine transformation from correctly matched triangles should transform other stars in the field to sky positions close to their catalog positions. This 
is typically used as a criterion for validating a star matching; a large percentage of stars in the field should be matched with the catalog.
Using the matched stars, we can solve for the mapping (the astrometric solution) between the pixel coordinate and the position in the plane of the sky as an affine transformation,
and thus the right ascension (RA) and declination (Dec). Because of non-ideal optics, we often need to go beyond the affine transformation
to use two-dimensional lower order polynomials to model the field distortions for more accurate 
astrometric solutions (see \cite{Zhai2018} for details).
For our TMF system with only a FOV of 6', an affine transformation is sufficient for 10 mas accuracy.
A 3rd order polynomial is needed to achieve 5 mas accuracy. For SRO1 and SRO2, a 3rd order polynomial 
is sufficient for achieving 50 mas astrometric solution. 

{\it Target processing} encompasses identifying a specific follow-up target or searching for new objects. In general, we first removed star signals to by setting pixels near detected stars to zero assuming we have estimated and subtracted the sky background \citep{Zhai2014}, so that we deal with frames with noises and signals from the target or objects to be detected. For a follow-up target with a known sky rate of motion, we can stack up images to track the target. We also apply a spatial kernel matching the PSF to improve S/N.
The target is located by finding the pixel that has the highest S/N in the expected region of the field. Sometimes, the target is too faint or the ephemeris
has uncertainties larger than what was estimated, human intervention is needed to help identify the object in the field.

In NEO search mode or when recovering follow-up targets with large uncertainties in ephemerides, 
we need to use GPUs to perform shift/add over a grid of velocities covering the rate of interest to detect
 signals above an S/N threshold, which we use 7.5 to avoid false positives \citep{Zhai2014}. 
The detected signals are clustered in a 4-d space (2-d position and 2-d velocity) to keep only the position and velocity with the highest S/N.
The last step in target processing is a least-squares fitting (moving PSF fitting) using a model PSF to fit the intensities of the moving object in the datacube to refine the positions and velocities of the detected signals \citep{Zhai2014, Zhai2018}.

\subsubsection{Search for NEOs using GPUs}
The advantage of ST for NEO search and recovery is the improved sensitivity from avoiding the trailing loss at the price of a large amount of
computation for processing the short exposure data cubes. For example, our SRO1 system typically uses a 5 second exposure time and integrates
100 frames. The camera frame size is 61 MPix giving a data cube of size about 12.2 GB stored in raw data as unsigned 16-bit integers. During data
processing, the data are stored as 32-bit floating point numbers, which means 24.4 GB of memory.
Our velocity range of interest is $\pm 0.63 as/sec$ (0.5 pixel/sec for a pixel scale of 1.26 as) over both RA and Dec and we use a 100$\times$100 grid to cover this range with a grid spacing of 0.0126 as/sec. This 
means that our maximum rate error is about $\pm 0.0063 as/sec$. For 500-second integration, the maximum streak length along a row or column due
to this rate error is about 3.2 as or 2.5 pixels. Since our best PSF has a FWHM of about 2.5 pixels, the trailing loss due to digital tracking error is less
than 12\% as shown in Fig.~\ref{fig:trailingLoss}. The amount of computation is $61\times 10^6 \times 100 \times 100 \times 100 \sim 6.1 \times 10^{13}$ FLOPS per data cube.

Fortunately, modern GPUs like a Tesla V100 allow us to process data in real-time; a single Tesla V100 with 32 GB memory can
perform the search in about 440 seconds. The performance is not limited by the GPU's processing speed but by the memory bandwidth, especially how to 
efficiently use the cache memory.
We note that the velocity grid spacing is determined so that the rate error due to discretization only causes a streak (in post-processing) of no more than 1 PSF per integration.
A typical velocity grid spacing is then 2 PSF per integration time. For our SRO1 system, the velocity grid spacing is 
$\sim 2 \times 2.5$pix/500 sec $\approx$0.01 pix/s. 
The velocity grid is $\pm$ 50 in both RA and Dec giving a range of rate of $\pm 0.63 as/s$.

When we recover an NEO whose ephemeris becomes highly uncertain, we need to search in the neighborhood of the 
expected location according to the ephemeris and to cover at least a region of the 3-$\sigma$
uncertainty of the ephemeris. The velocity grid to search should cover around the  projected rate of motion
also covering at least the 3-$\sigma$ uncertainty of the rate. We use SRO2 to do NEO recovery.
Because 0.63 as/s $\approx$15 deg/day,
for a newly discovered object without follow-up for 1-2 nights, we typically can recover these objects without
trouble because typically the position errors are less than 10 deg and rate errors are less than 10 deg/day. Since SRO2 has a FOV size of 2.6 deg $\times$ 1.7 deg and 
for recovery, the uncertainties is along the track of the NEO, thus we only need a one-dimension (instead of 2-d) search, so the computation load
is much less for recovering an object than the general NEO search.

We note that the range of rate for searching a moving object is only limited by the total amount of computation needed. With multiple GPUs, it is possible to search over even
larger range of rate to detect for example earth orbiting objects.

\subsubsection{Reduce systematic astrometric errors}
Gaia's unprecedented accuracy allows us to push NEO astrometry to 10 mas \citep{Zhai2018}. Highly accurate astrometry requires properly handling systematic astrometric errors such as the DCR effect, star confusion, and timing error.

\subsubsubsection{Difference Chromatic Refraction (DCR) Effect}
Unless observing at the zenith, the light rays detected are bent by the atmosphere due to refraction. 
Because the index of air refraction depends on the wavelength of light $\sim 1/\lambda^2$ \citep{Ciddor2002}, the atmospheric refraction bend more the blue light than red light.
This introduces a systematic error in astrometry, the DCR effect, if the target and reference objects have different spectra.
If a narrowband filter is applied, the DCR effect becomes much less because the variation of atmospheric refraction is significantly reduced by limited passband.
However, to detect as much photon as possible to improve S/N, we typically use broadband or clear filters.

DCR effects can be modeled using an air refraction model \citep{Ciddor2002} and the spectra of target and reference objects \citep{Stone1996}.
For 10 mas accuracy, we found it sufficient to use a simple empirical model based on color defined as difference of Gaia magnitudes in blue and red passbands \citep{Andrae2018} as discussed in appendix~\ref{sec:dcr}.
The DCR correction in RA and Dec between reference color $C_{\rm ref} = (B-R)_{\rm ref}$ and target color $C_{\rm tar} = (B - R)_{\rm tar}$ is expressed as
 \beq
    DCR = \tan(\theta_z) (\sin \phi_z, \cos \phi_z) \left [(a C_{\rm ref} + b C_{\rm ref}^2) - (a C_{\rm tar} + b C_{\rm tar}^2 ) \right ] \,,
\label{dcr_model} 
 \eeq
where $\theta_z$ is the zenith angle, complementary to the elevation angle, $\phi_z$ is the parallactic angle
between the zenith and celestial pole from the center of the field.
In general, we do not have spectral information of NEO candidates from the NEOCP, 
so we assume a solar spectrum for them assuming they reflect sun light uniformly across the band as a leading order approximation with $C_{\rm tar} \approx 0.85$ (estimated using Fig. 3 in \cite{Andrae2018} assuming an effective temperature of 5800 K for solar spectrum).
 
\begin{figure}[h]
\epsscale{1.1}
\plottwo{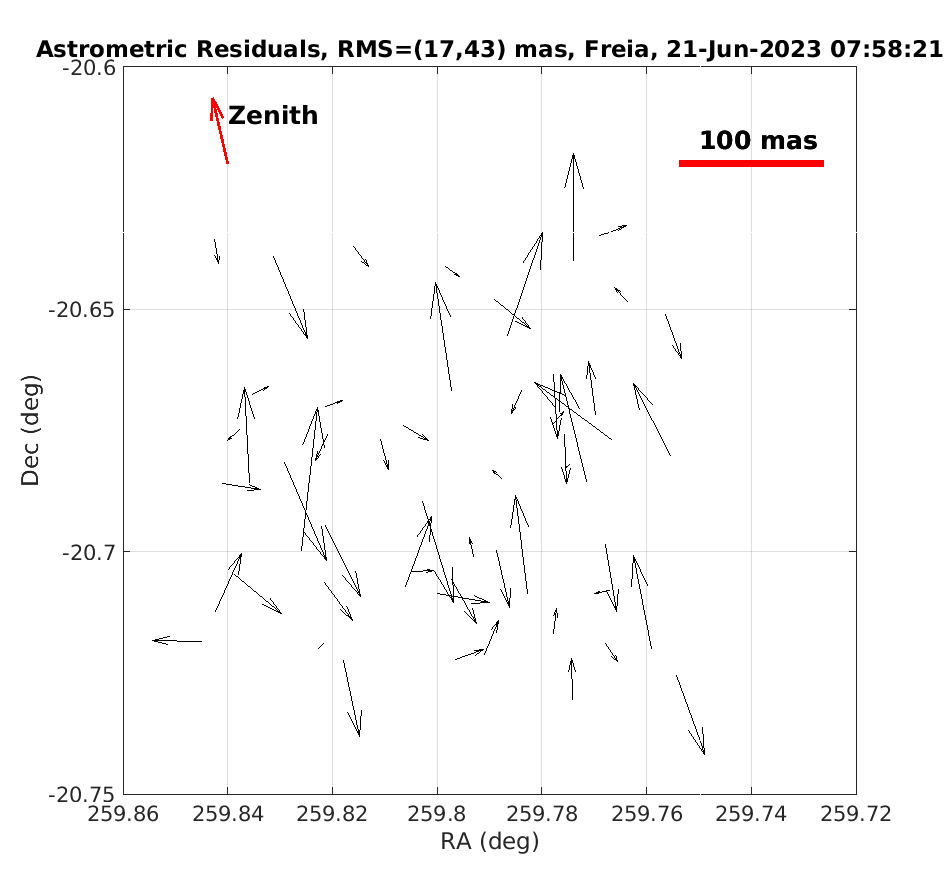}{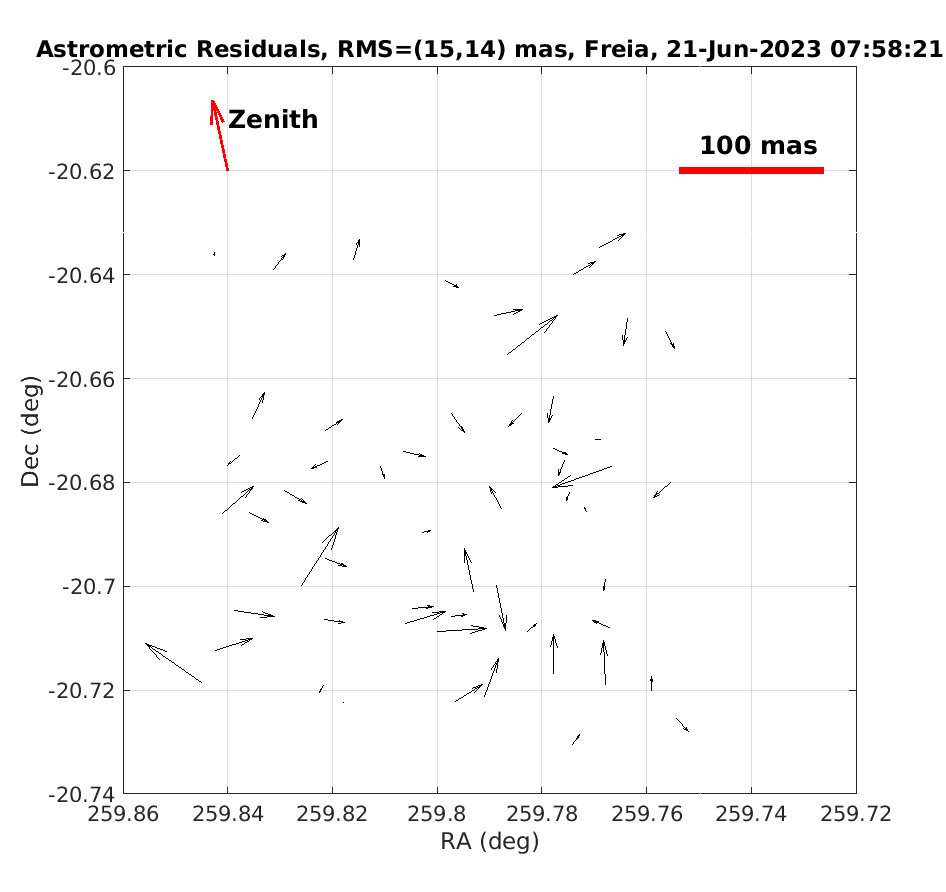}
\caption{Astrometric solution residuals using a 3rd order polynomial to fit field distortions with the Gaia DR2 catalog to show differential chromatic refraction (DCR) effect with a clear filter.
Left plot shows residuals without DCR correct and right plot shows residuals after correcting DCR effect using a simple quadratic color model. \label{fig:dcr_off_on}}
\end{figure}

\begin{figure}[h]
\epsscale{0.5}
\plotone{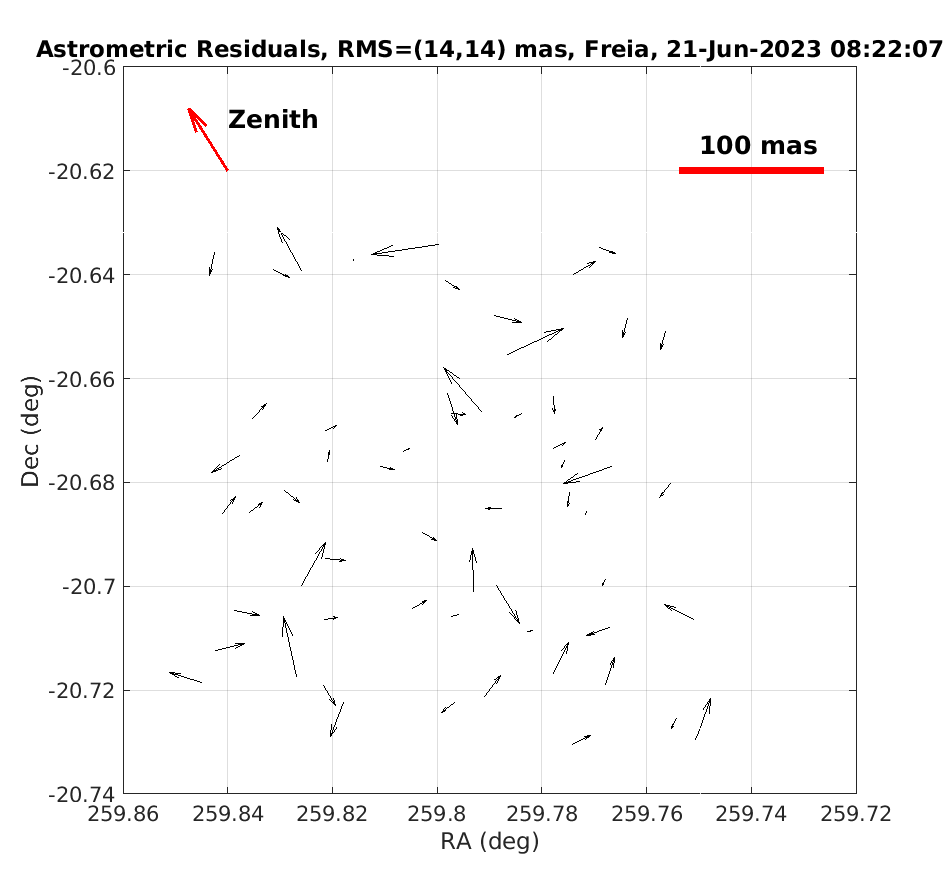}
\caption{Astrometric residuals of a field observing asteroid Freia (76) using an i-band filter the DCR effects are too small to identify.\label{fig:astr_sol_i}}
\end{figure}

\begin{figure}[h]
\epsscale{0.5}
\plotone{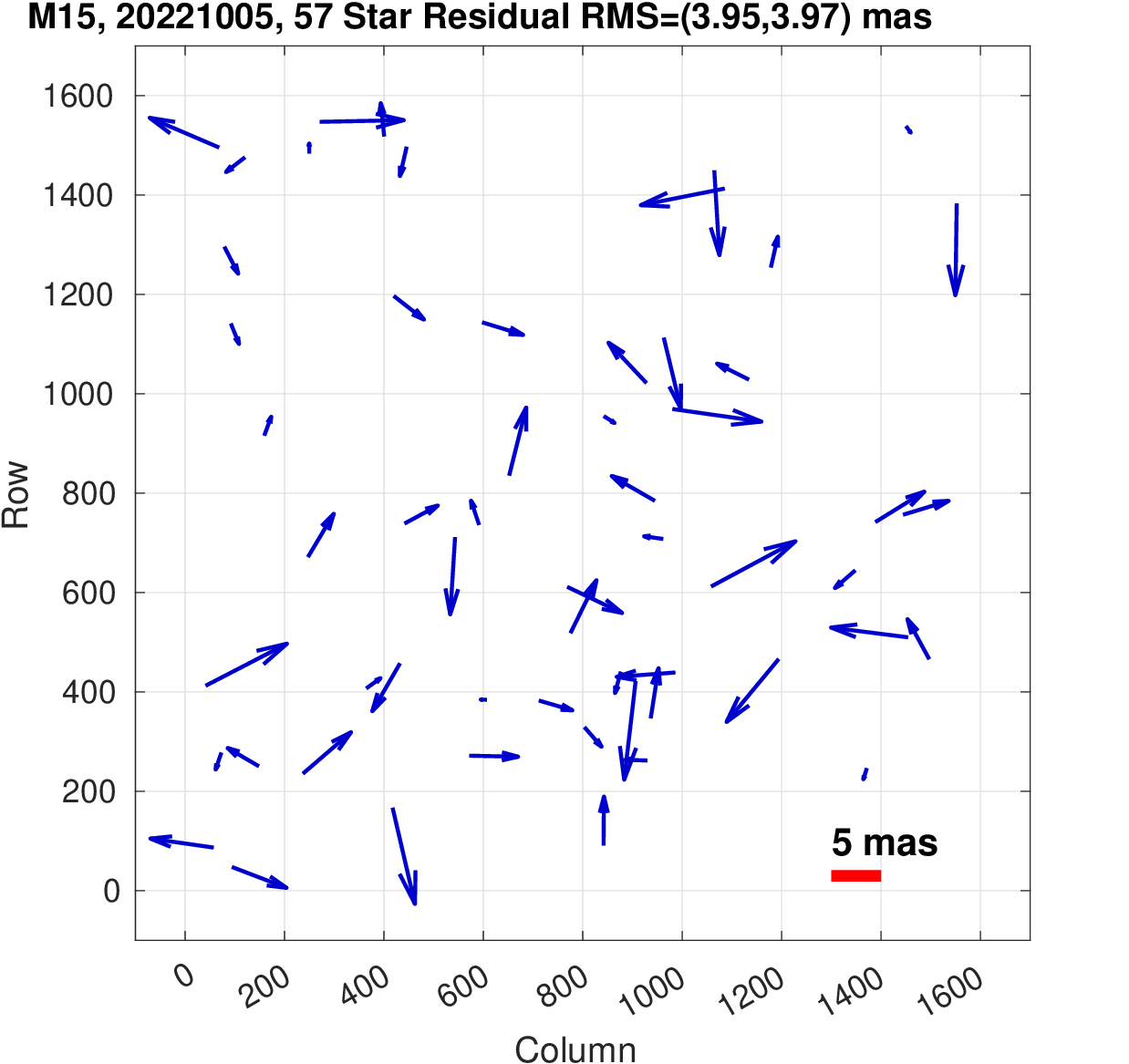}
\caption{Astrometric residuals after correcting DCR effects using a clear filter, TMO (654), showing consistency with Gaia DR2 an RMS of 5 mas using a third order polynomial field distortion model. \label{fig:astr_sol}}
\end{figure}

As an example, when we use a clear filter, without any DCR correction, the astrometric residuals for a field observing Feria (76)
with a clear filter are displayed in the left plot in
Fig.~\ref{fig:dcr_off_on}, where the dominant astrometric residuals are along the direction of zenith from the field.
Using Eq.~ (\ref{dcr_model}) to correct the DCR effect for a target with solar spectral type,
 we significantly reduced the RMS of the residuals from more than 40 mas to about 15 mas and the directions of residuals appear random.
As a comparison, if we apply a Sloan i-band filter (700-800 nm), the DCR effect becomes smaller than 10 mas because of the limited bandwidth
and the less color dependency for longer wavelength because of the dependency of refraction index
$\sim1/\lambda^2$  on wavelength. An additional limitation of bandwidth comes from the falling sensitivity of the Photometrics camera at the longer wavelength making
an effectively narrower passband shift to the 700 nm side of the passband. Indeed, without DCR correction, we found that for the 
same field observing Feria, the RMS of astrometric residuals is about 14 mas as shown in Fig.~\ref{fig:astr_sol_i}.
Note that the residuals shown in Fig~\ref{fig:astr_sol_i} and the right plot in Fig.~\ref{fig:dcr_off_on} contain significant 
photon noises. Fig.~\ref{fig:astr_sol} displays residuals observing M15 where only the residuals of 
bright stars in the field are displayed. Here residuals are not limited by photon noises and we are able to achieve better than 5 mas accuracy.

\subsubsubsection{Target position estimation and confusion elimination\label{sec:confusion}}
Star confusion is another source of systematic errors for astrometry. We exclude the frames where the 
NEO gets close to a star whose light could affect the centroiding of the NEO. The centroiding error due to
star confusion in units of the FWHM of the PSF is estimated as the gradient of
the intensity of the star at the NEO's location relative to the NEO peak intensity divided by its FWHM
$\sim |grad(I_{\rm star}(at \; NEO))|/ \left ( I_{\rm NEO} (peak) / FWHM \right) $. Based on this estimation,
we exclude frames that could lead to centroiding errors above a threshold, {\it e.g.} 0.01, which should be determined by
accuracy requirement assuming this corresponds to a centroiding error $\sim$ 0.01 FWHM.
The ``moving-PSF'' fitting is done for frames without confusion.
The general cost function for the least-squares fitting of ``moving PSF'' to the whole data cube
\beq
  C(v_x, v_y, x_c, y_c, \alpha, I_0 ) \equiv \sum_{t=1}^{N_f} \sum_{x,y} \left | I (x, y, t) -\alpha P \left (x {-}X(t), y {-}Y(t) \right )  - I_0 \right |^2 w(x, y, t)
\eeq
has a weighting function $w(x,y,t)$, which can be set to 0 for frames with confusion. $P(x,y)$ is the PSF function, typically a Gaussian PSF,
and $\left (X(t), Y(t) \right )$ represent the location of the object in frame $t$, and $N_f$ is the total number of frames.
To minimize the variance of the estimation, the weight can be chosen to be the inverse of the variance of
the measured $I(x,y,t)$, including photon shot noise and sky background, dark current and read noise 
according to the Gauss-Markov theorem \citep{Luenberger1969}. For frames without 
confusion, we usually choose $w=1$ for simplicity because the noise in $I(x,y,t)$ 
is not the limiting factor of accuracy for most of our targets.
For vast majority of objects, the motion can be modeled as linear:
\beq
   X(t) = x_c + v_x \left (t - (N{+}1)/2 \right ) + \epsilon_x (t) \,, Y(t) = y_c + v_y \left (t - (N{+}1)/2 \right ) + \epsilon_y (t)
\eeq
where $(x_c, y_c)$ is the object position at the center of the integration time interval and $(v_x, v_y)$ is the velocity.
$(\epsilon_x (t), \epsilon_y(t))$ are the residual tracking errors (fractions of a pixel) with respect to sidereal 
after re-registering frames by shifting an integer amount of pixels satisfying $\sum_{t=0}^N (\epsilon_x(t), \epsilon_y(t)) = (0, 0)$.
$\epsilon_{x,y}(t)$ can be estimated as the averages of the centroids of reference stars in each frame after frame re-registration (integer shifts of frames).
The location $(x_c, y_c)$, velocity $(v_x, v_y)$, and parameters $\alpha$ and $I_0$ are solved simultaneously using a least-squares fitting.

\subsubsubsection{Accurate timing}
Accurate timing is crucial for generating precision astrometry.
The timestamp for a camera frame in general should correspond to the epoch at the center of the exposure time window for the frame.
If possible, hardware timing with a GPS clock is desired because software timestamps 
from non-real-time operating systems can have errors due to indeterministic runtime behaviors.
Hardware timing can be achieved by using a GPS clock to trigger the start of the exposure at a preset time
or by letting a GPS clock record the signals generated by a camera upon the completion of a frame.
For example, our Photometrics Prime 95B used at TMF can accept a trigger signal 
from a Meinberg GPS clock to initiate the exposure of frames.
Camera reference manual should be referred to interpret timestamps correctly. 
For example, CMOS cameras have both global shutter and rolling shutter modes and there could be dead time between frames.
We usually operate CMOS cameras in the rolling shutter mode for high frame rate and low read noise.
Rolling shutter mode delays the exposure time window of each row of the image by a small constant time offset relative to the previous row in the order of readout.
It is important to account for this small time delay between the consecutive rows because we usually have the timestamps for reading out the first or last row, but the target is
observed at some row in between. This delay is 19.6 usec per row for our Photometrics camera at TMF. 
A useful test for understanding the details of timing is to use a GPS clock to trigger both the camera and an LED light
and examine the recorded frames. Using this test, we found our Photometrics camera has an extra delay of 50 msec 
for the first frame to start after the trigger signal for the camera. 

When excluding frames to avoid star confusion, we need to derive
astrometry based on the timestamps of the frames that do not have confusion.
For our TMF system, we are confident that our timing accuracy is better than 10 msec,
which was confirmed by the small ($<$ 0.1 as) astrometric residuals from observing a GPS satellite (C11)
relative to the ephemerides from Project Pluto (https://www.projectpluto.com/) 
and the results from the International Asteroid Warning Network (IAWN) 2019 XS timing campaign\citep{IAWN2021} that we participated in.

\section{Results}

In this section, we present results from our instrument on the Pomona 40-inch telescope at the TMF and robotic telescopes at the SRO. 
Our instrument at TMF (654) has consistently produced accurate astrometric measurements and our robotic telescopes at SRO 
has been able to detect faint NEOs at about mag = 20.5.

\subsection{Astrometric Precision using Synthetic Tracking}
\label{sec:centAccuracy}
\begin{figure}[h]
\epsscale{0.95}
\plottwo{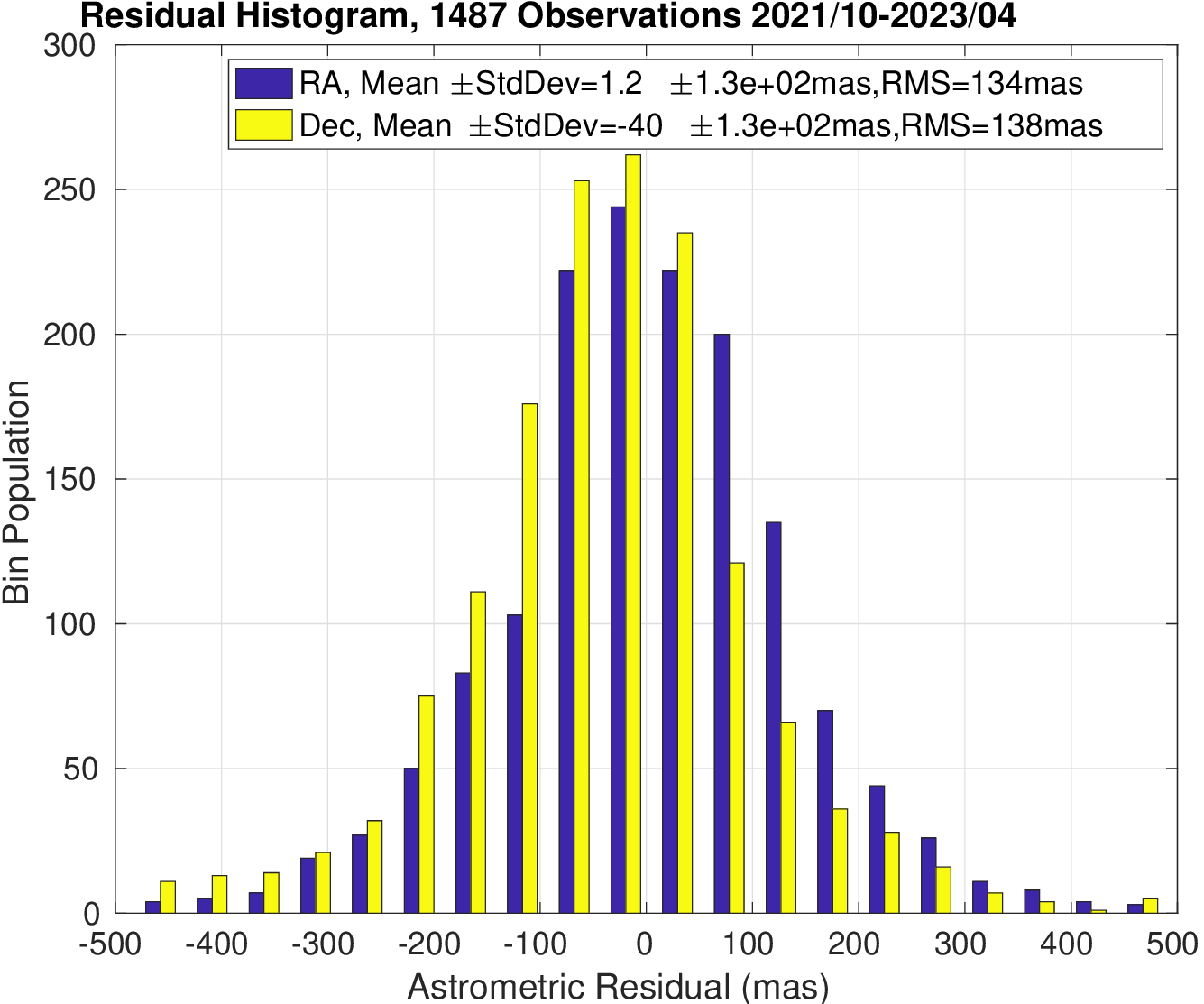}{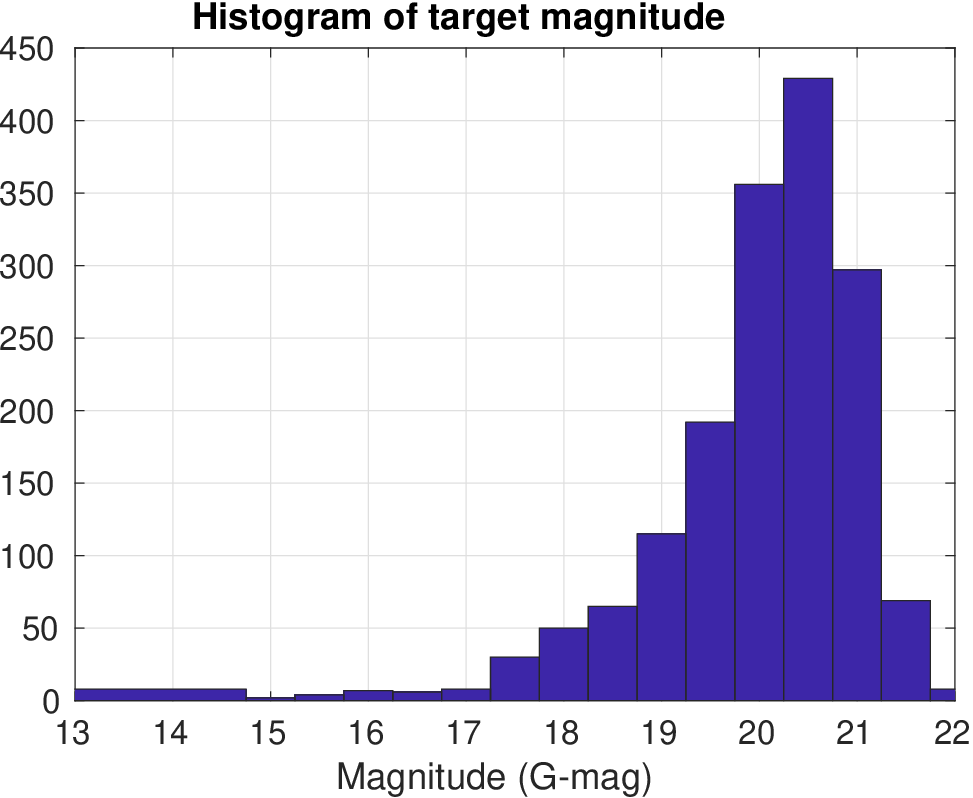}
\caption{Histogram of residuals (left) and target magnitudes.\label{fig:res_mag_hist}}.
\end{figure}

Synthetic tracking avoids streaked images by having exposures short enough 
so that the moving object does not streak in individual images,
allowing us to achieve NEO astrometry with accuracy similar to stellar astrometry.
We have been regularly observing NEOs from the MPC confirmation page since 2021 with a support by the NEOO
program targeting NEOs brighter than 22 mag.
The left and right plots in Fig.~\ref{fig:res_mag_hist} respectively display 
NEO astrometric residuals from subtracting the JPL Horizons ephemerides and the estimated apparent NEO magnitudes 
from Oct 2021 to Apr 2023 after we fixed a timing error.

\begin{figure}[h]
\epsscale{0.95}
\plotone{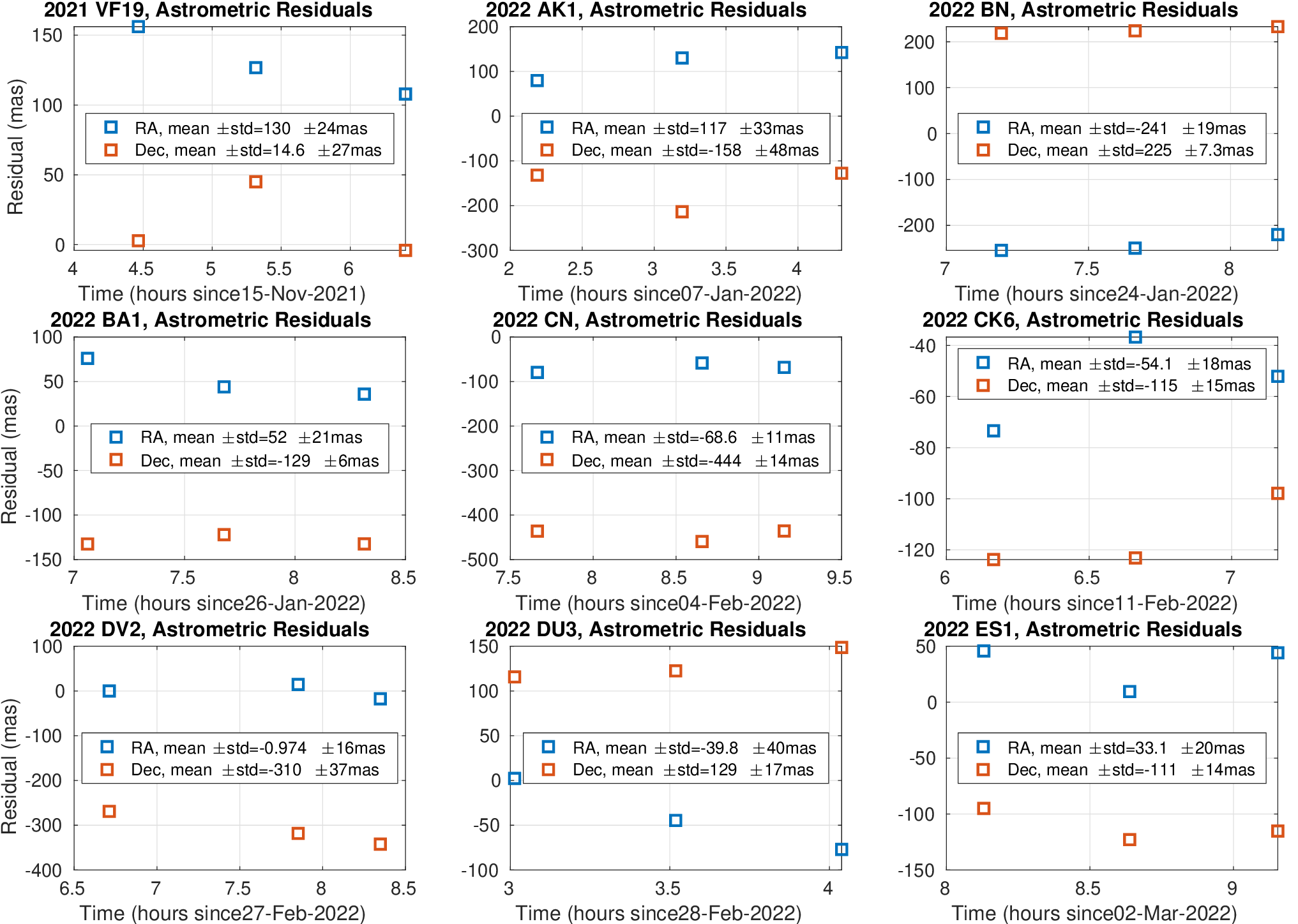}
\caption{Examples of uncertainties of NEO ephemerides.\label{fig:eph_uncertain}}.
\end{figure}

\begin{figure}[h]
\epsscale{0.95}
\plottwo{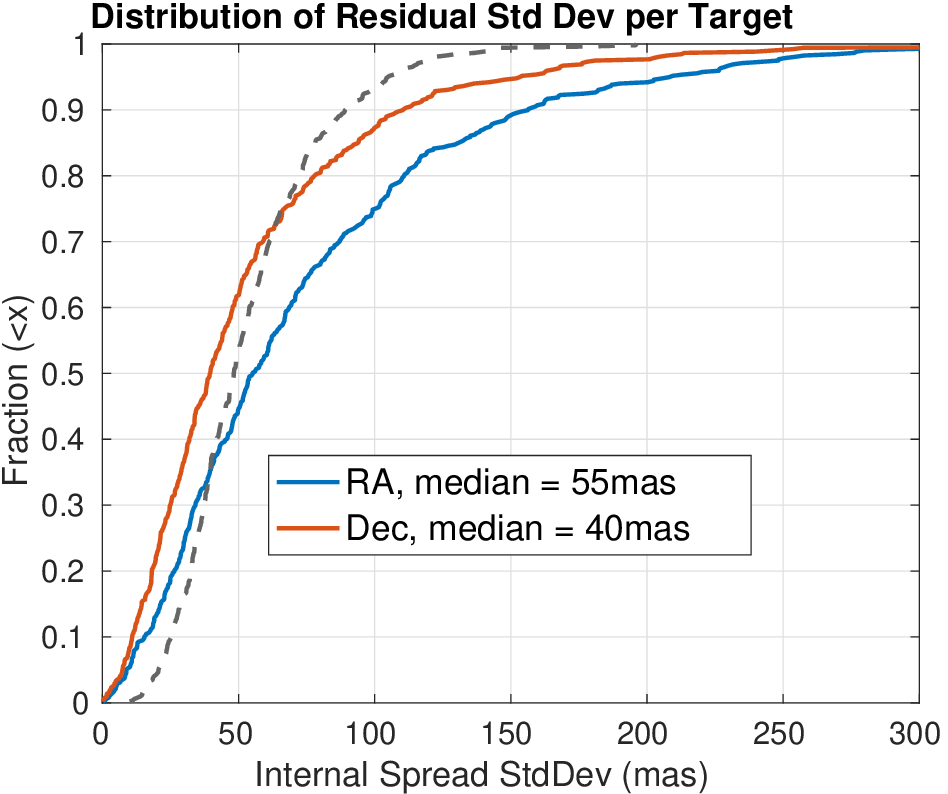}{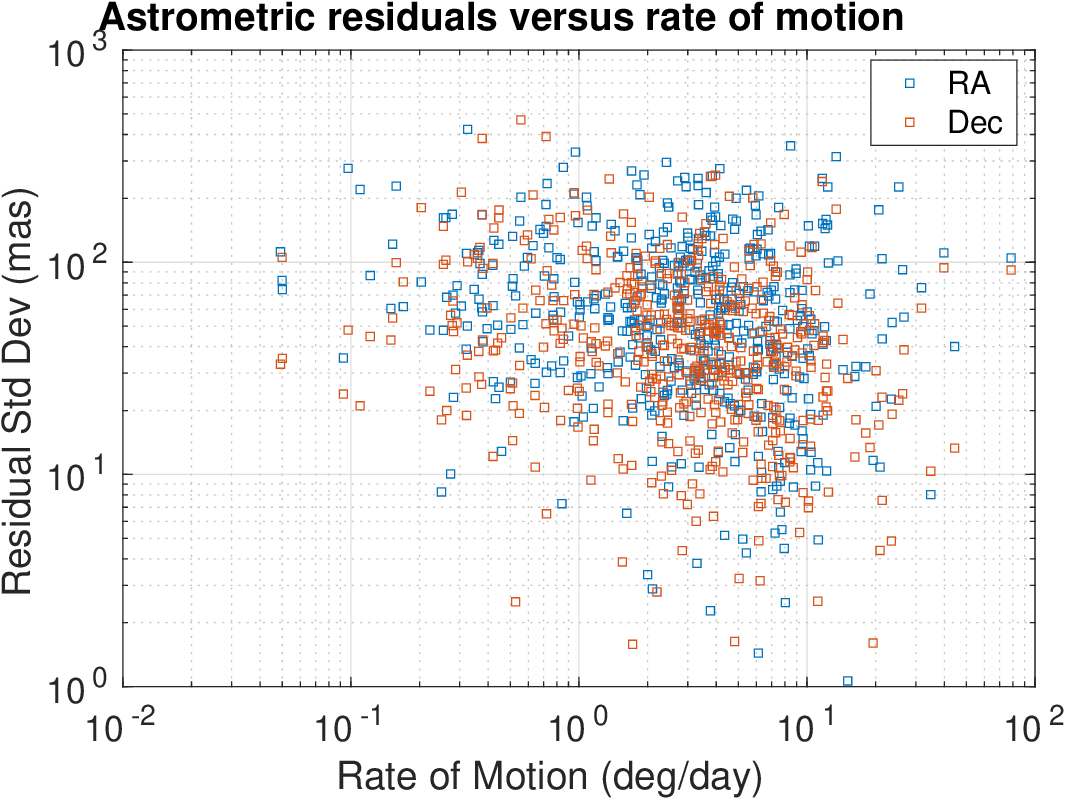}
\caption{Dstribution of standard deviations of the measurements for each target (right), where the black dashed line represent an estimated value of the 1-$\sigma$ uncertainty
of the astrometric measurements (left) and the dependency of residual standard deviations versus rate of motion (right).\label{fig:std_dist_rate}}.
\end{figure}

\begin{figure}[ht]
\epsscale{0.6}
\plotone{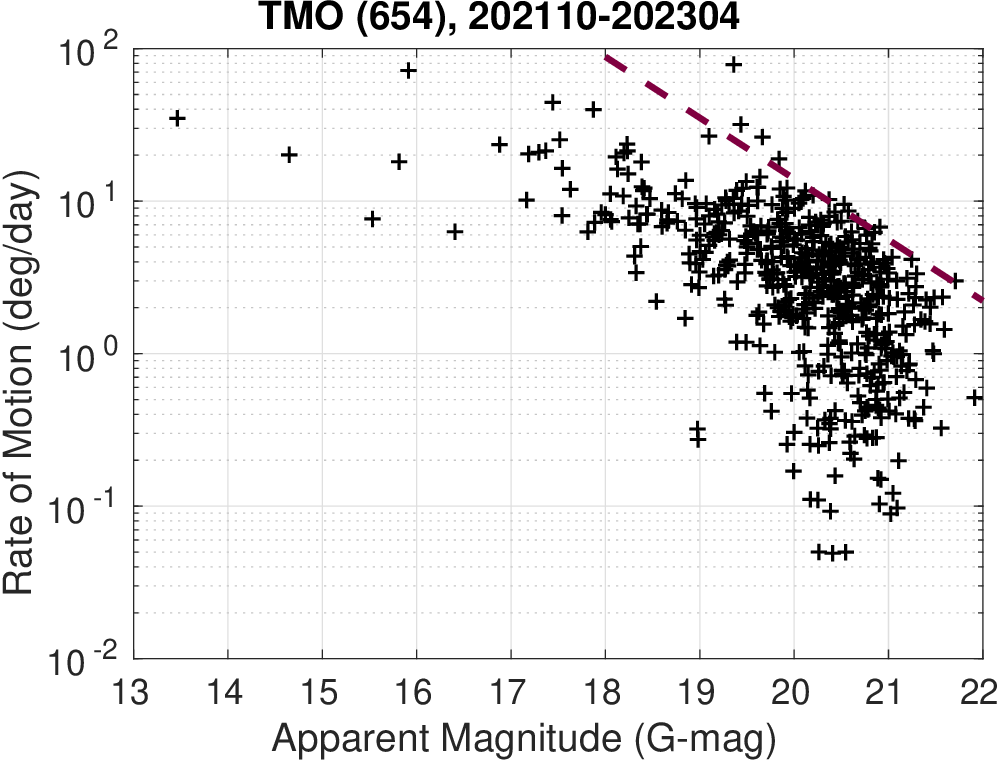}
\caption{Rate of motion versus apparent magnitudes of the NEOs from the confirmation page during 2021/10 to 2022/04.\label{fig:rate_mag}}
\end{figure}

While the RMS (root-mean-squares) of our residuals are about 130 mas, which is among the best category of NEO astrometry accuracy (Veres 2017), we believe our accuracy is better than 130 mas
because these residuals include the uncertainties of the ephemerides of the NEOs, which are mostly new discoveries with observations covering only a short arc of the orbit.
To illustrate this, Fig.~\ref{fig:eph_uncertain} shows residuals of nine NEOs, where we can clearly see that residuals are biased and the spread of our measurements are much smaller.
These biases are consistent with JPL Horizons 3-$\sigma$ uncertainty estimates as shown in Table~\ref{table:2}.
\begin{table}[ht]
\caption{JPL Horizons NEO Ephemeris Uncertainties and the Residual Biases\label{table:2}}
\centering
\begin{tabular}{|c|c|c|c|c|c|}
\hline
\textbf{Designation} &  \textbf{Epoch} & \textbf{3$\sigma$ RA (as) } & \textbf{3$\sigma$ Dec (as)} & Res RA (as) & Res Dec (as) \\
\hline
2021 VF19  & 20211115UT05:00 &0.532      &  0.533 & 0.13 & 0.015 \\
\hline
2022 AK1 &20220107UT03:00&  0.712    &  0.478 & 0.12 & -0.16\\
\hline
2022 BN & 20220124UT07:30 & 0.517  &    0.535 & -0.24 & 0.23 \\
\hline
2022 BA1 & 20220126UT08:00 &0.204  &    0.218 & 0.052 & -0.13\\
\hline
2022 CN & 20220204UT08:30 & 0.469   &   0.455 & -0.069 & -0.44\\
\hline
2022 CK6 & 20220211UT06:30& 0.531  &    0.416 & 0.054 & -0.12 \\
\hline
2022 DV2 & 20220227UT07:30 & 0.448  &    0.852 & 0.001 & 0.31 \\
\hline
2022 DU3& 20220228UT03:30 & 0.168   &   0.223 & 0.040 & 0.13\\
\hline
2022 ES1 &20220302UT08:30 & 0.227   &   0.226 & 0.033 & -0.11 \\
\hline
\end{tabular}
\end{table}
If instead, we use the standard deviations of our measurements for each target as a measure of astrometric uncertainties 
to remove the uncertainties in the ephemerides, we get the
distribution shown in the left plot in Fig.~\ref{fig:std_dist_rate}. The right plot shows the residuals with the rate of motion. A slight trend of going down is due to the fact that highly moving
objects tend to be brighter because of the trailing loss in survey detections as shown in Fig.~\ref{fig:rate_mag}.
\clearpage

\begin{figure}[ht]
\epsscale{0.75}
\plotone{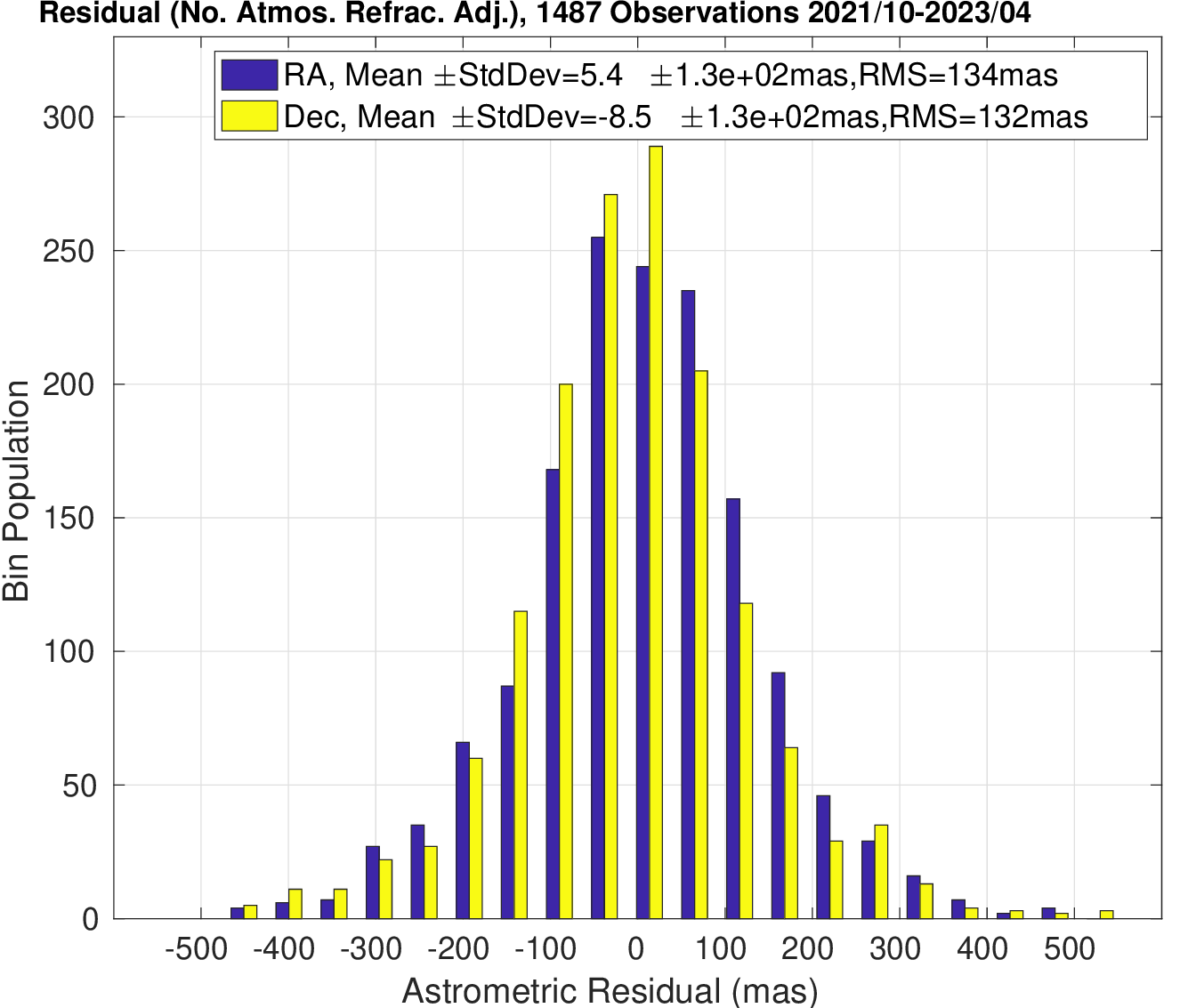}
\caption{Histogram of residuals without correcting the differential chromatic atmospheric refraction effect.\label{fig:res_no_color_corr}}.
\end{figure}

We note that the bias of about $\sim$-40 mas in Dec seems to be significant because the mean residual is derived from observations about 500 NEOs
with expected random standard error of only $130 / \sqrt{500} \sim 5.8$ mas.
This bias could be related to the fact that we correct the atmospheric DCR effect.
If we remove the DCR corrections from our astrometry, we found the bias is close to zeros as shown in Fig.~\ref{fig:res_no_color_corr}. 

\begin{figure}[h]
\epsscale{0.75}
\plotone{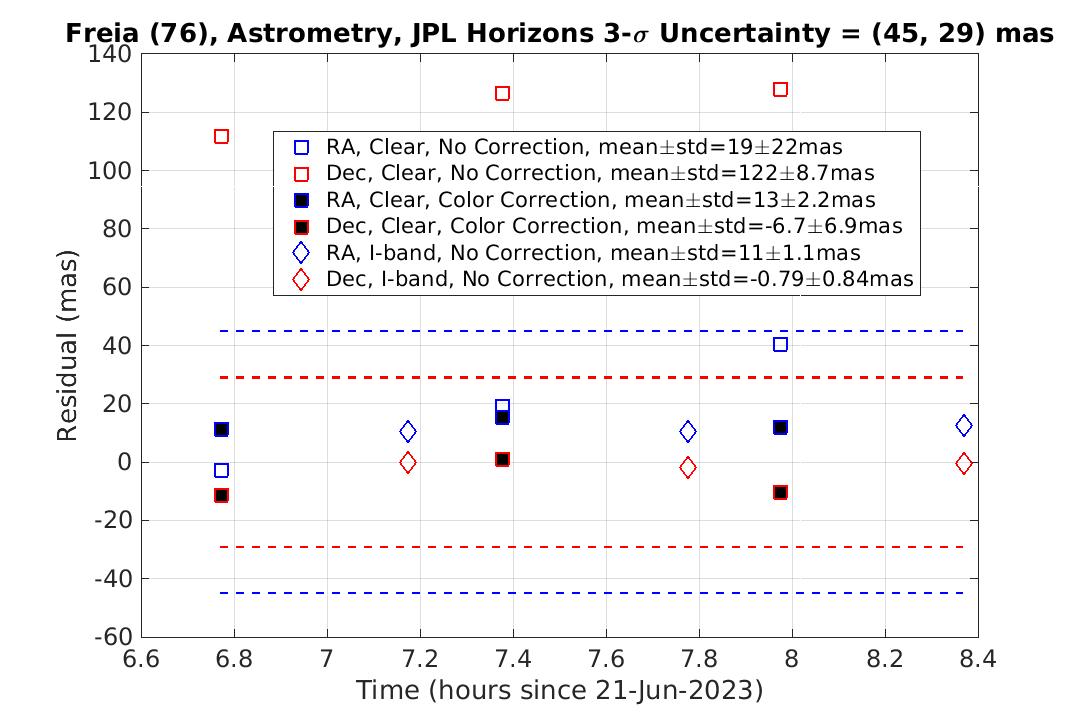}
\caption{An example of DCR correct for clear filter. \label{fig:dcr_corr}}
\end{figure}

The DCR correction is necessary for achieving 10 mas astrometry consistency with the
Gaia DR2 catalog as shown in Fig.~\ref{fig:dcr_off_on} and Fig.~\ref{fig:astr_sol} when using a clear filter.
We also found that the DCR correction would give consistent astrometry for bright NEAs like Freia (76). 
Fig.~\ref{fig:dcr_corr} displays residuals of astrometry of our observations on target Freia (76) subtracting the JPL Horizons ephemerides.
The squares represent observations using a clear filter. The solid (empty) squares represent results results with (without) DCR corrections for the clear filter.
The diamonds are the observations of Freia using a Sloan i-filter, which makes DCR effect significantly smaller due to its limited bandwidth, the CMOS chip response, and the longer wavelength in view of the atmospheric refraction dependency on wavelength  $\sim \lambda^{-2}$, which is confirmed
by the low residuals of astrometric solution without any DCR correction shown in Fig.~\ref{fig:astr_sol_i}.

The fact that the DCR correction makes broadband astrometry much closer to an i-band filter with lower astrometric residuals 
as shown in the right plot in Fig.~\ref{fig:dcr_off_on} validates our approach of DCR correction.
We assume solar spectrum for NEOs by default. However, NEO spectra can deviate from the solar spectrum,
so improvements can be made by measuring the color of the NEOs.
To our knowledge, vast majority of the NEO observations do not correct the DCR effects,
therefore it is possible for these astrometric measurements to be biased due to DCR effects
depending on average elevations of observations; measurements with DCR corrections like ours only appear to be biased along Dec because DCR effects along RA can be positive or negative depending on the sign of hour angles, thus not introducing an overall bias.
It would be interesting to further understand this bias by comparing all the observations with and without DCR corrections
and the filters applied.

\subsection{Detection Sensitvity}
Using ST to survey fast-moving NEOs avoids the trailing loss, and thus reaches detection sensitivity as if we were tracking the targets.
This enables small telescopes to observe and detect NEOs using long integrations, which would not be possible if the trailing loss degrades the S/N as we integrate long (see Fig.~\ref{fig:sens_trade} for the degraded sensitivity when $\Delta t$ increases beyond the optimal exposure time).
We have been experimenting with our 11-inch telescope system SRO1 (U68) to survey NEOs with a 5-second exposure time and an integration of
100 frames, giving a limiting magnitude of $\sim$ 20.5 for clear dark nights.

Currently, our operation observes each field twice during the same night with about 45 minutes between the visit and revisit. With three telescopes (total 22 sqdeg FOV),
we can cover approximately 300 sqdeg per night and the whole sky in about 10 days.
Each ST observation gives an estimate of the sky position and rate of motion along RA and Dec for the target. 
We can determine whether two observations at different epochs are for the same object by checking 
whether the rates of motion from the two observations are consistent with the position changes between
the two epochs for a linear motion.
If the visit and revisit provide two consistent detections of the same object, we essentially have a
tracklet of four observations for the same object, thus the detections are reliable and 
the data is then reported to the Minor Planet Center (MPC).

\begin{figure}[h]
\epsscale{0.465}
\plotone{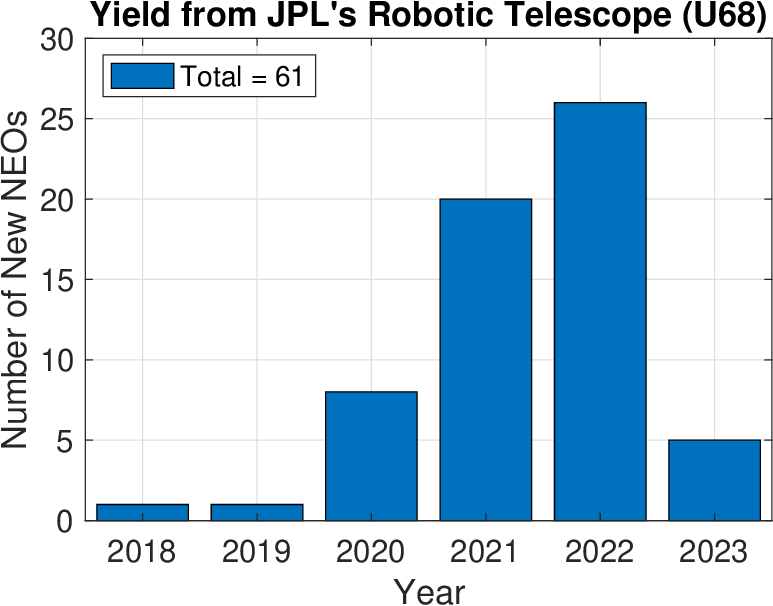}
\epsscale{0.44}
\plotone{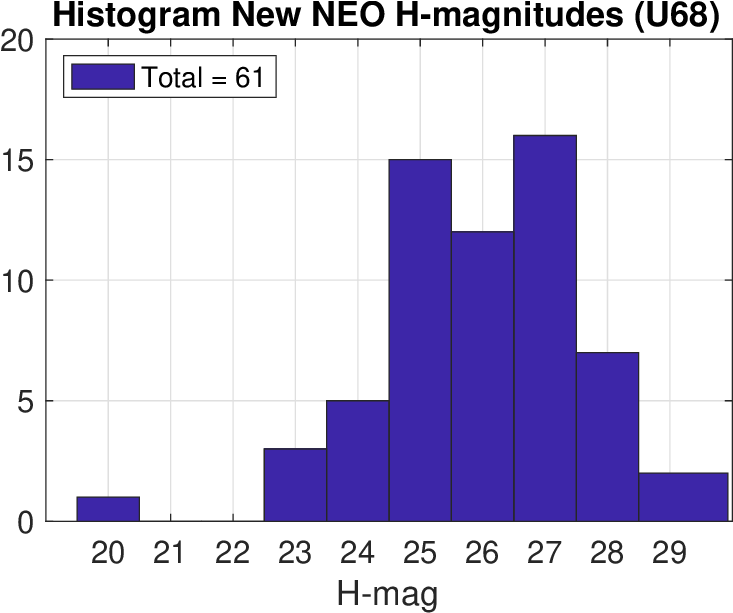}
\caption{Histogram of discoveries by JPL's robotic telescope at SRO (U68).\label{fig:U68_disc_stat}}.
\end{figure}

\begin{figure}[ht]
\epsscale{0.55}
\plotone{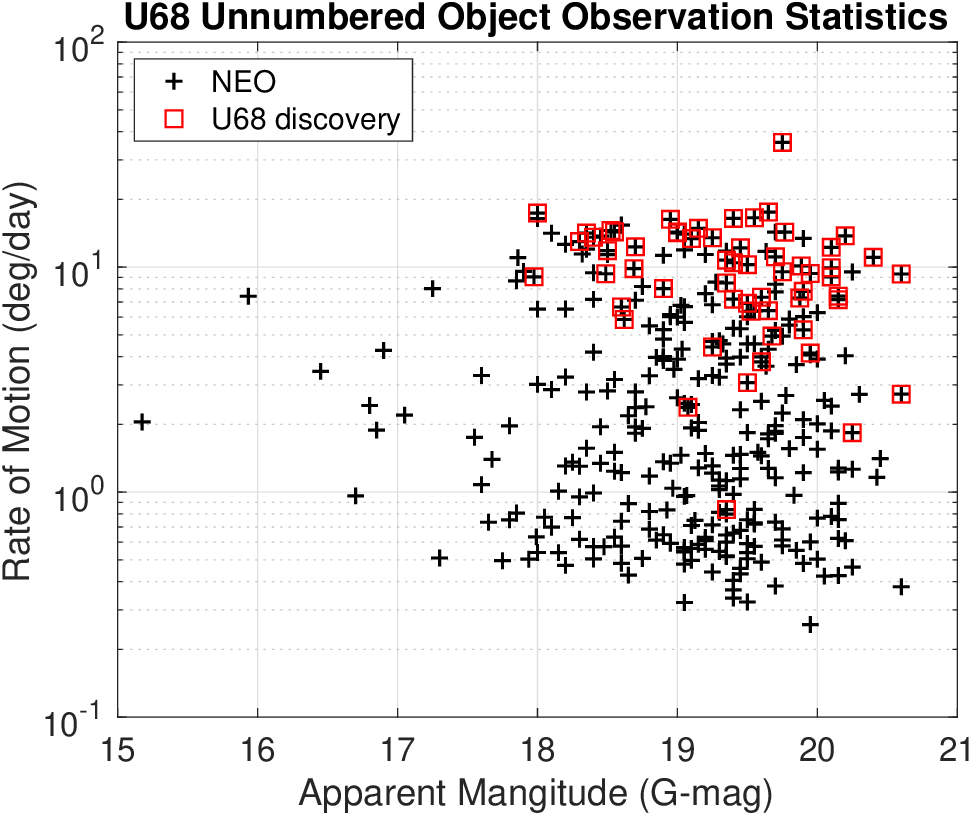}
\caption{The rate of motion and apparent magnitude plot for observations from JPL's robotic telescope at SRO (U68).\label{fig:U68_rate_mag}}.
\end{figure}

So far, we have discovered 61 new NEOs since 2018. 
A steady process has been made toward more efficient operation procedures, better imaging quality, and more reliable software 
as reflected by the number of detections shown in the left chart in Fig.~\ref{fig:U68_disc_stat}.
We are also working on an automated relay between the survey (SRO1) and the follow-up (SRO2) systems.
Fig.~\ref{fig:U68_rate_mag} shows the rate of motion and the apparent magnitudes for all the observations (black plus sign) that we reported to MPC.
The plus signs surrounded by red boxes are our new discoveries. 
These 11-inch telescopes can observe NEOs beyond 20.5 magnitude, regardless of the rate of motion, as we estimated in section~\ref{sec:sci_obs}.
This is different from what was shown in Fig.~\ref{fig:rate_mag}, where we clearly see a deficit 
beyond the dashed line (rate of motion = (14 deg/day) $\times 10^{-0.4(mag -20)}$, inverse with the brightness),
 showing clearly major survey facilities suffers trailing loss with detection magnitude inversely proportional to the rate of motion as shown in Fig.~\ref{fig:trailingLoss}.
Fig.~\ref{fig:U68_rate_mag} demonstrates the efficacy of ST in searching faint fast-moving NEOs complimentary to the major survey facilities.

\begin{figure}[ht]
\epsscale{0.48}
\plotone{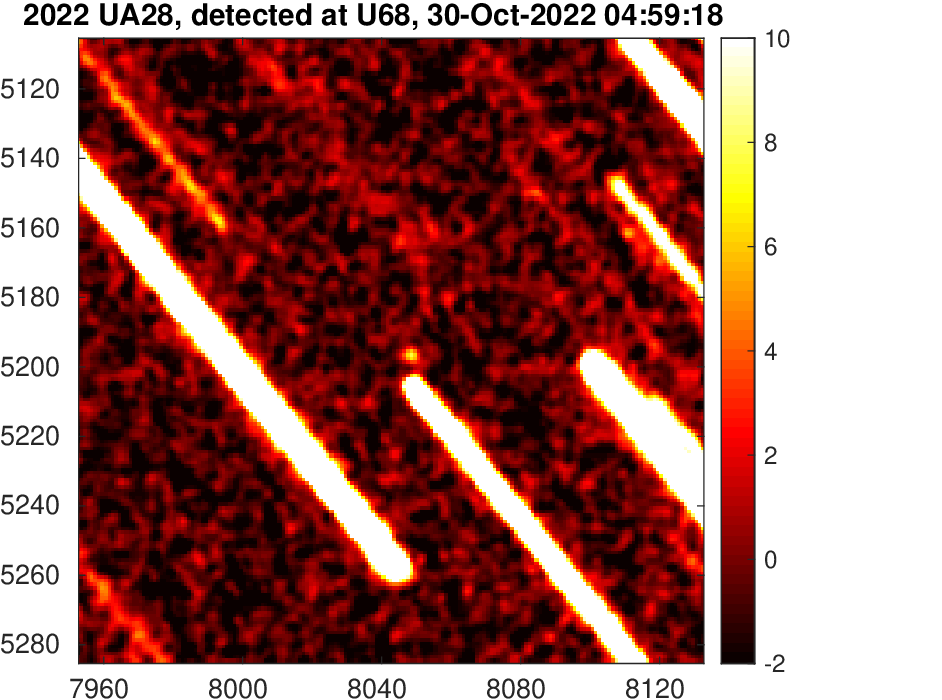}
\epsscale{0.48}
\plotone{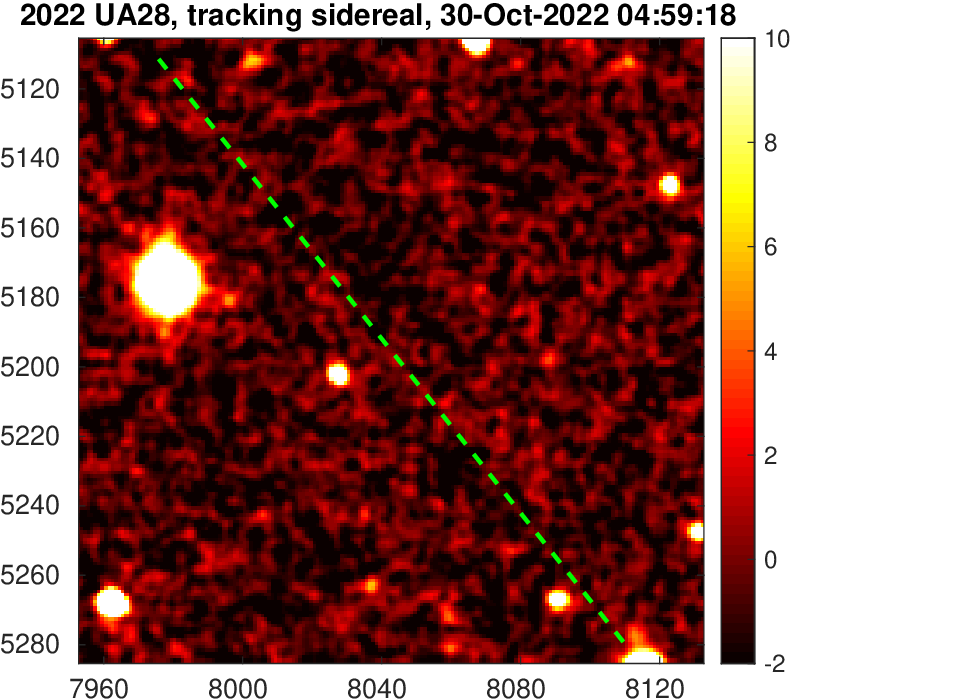}
\caption{An example of NEO (2022 UA28) discovered by JPL's robotic telescope at SRO (U68) Left image shows integration tracking 2022 UA28 (at the center) and right image is the integration tracking sidereal
with the NEO track marked as the green dashed line, where the trailing loss makes NEO signal buried in noises.\label{fig:2022UA28}}.
\end{figure}

Fig~\ref{fig:2022UA28} shows the detection of a fast-moving object, 2022 UA28, which represents the current frontier of faint fast-moving detection. The object was moving at a rate of 11 deg/day, (-7.6, 8.1) deg/day along (RA, Dec) with apparent magnitude of 20.5 mag. If using an exposure time of 30 s, the streak length would be 13.8 arcsec.
Assuming 2 as FWHM for PSF, the trailing loss would be a factor $\sim13.8/(1.5*2) ~ 4.6$, so more than 1.5 stellar magnitude fainter. (For PanSTARRS, this would even more as
the integration time is 45 s and the PSF is more compact than 2 as). We detected this object as an SNR about 8.3 enabled by ST.

\subsection{Robustness against Star Confusion}
\begin{figure}[ht]
\epsscale{0.30}
\plotone{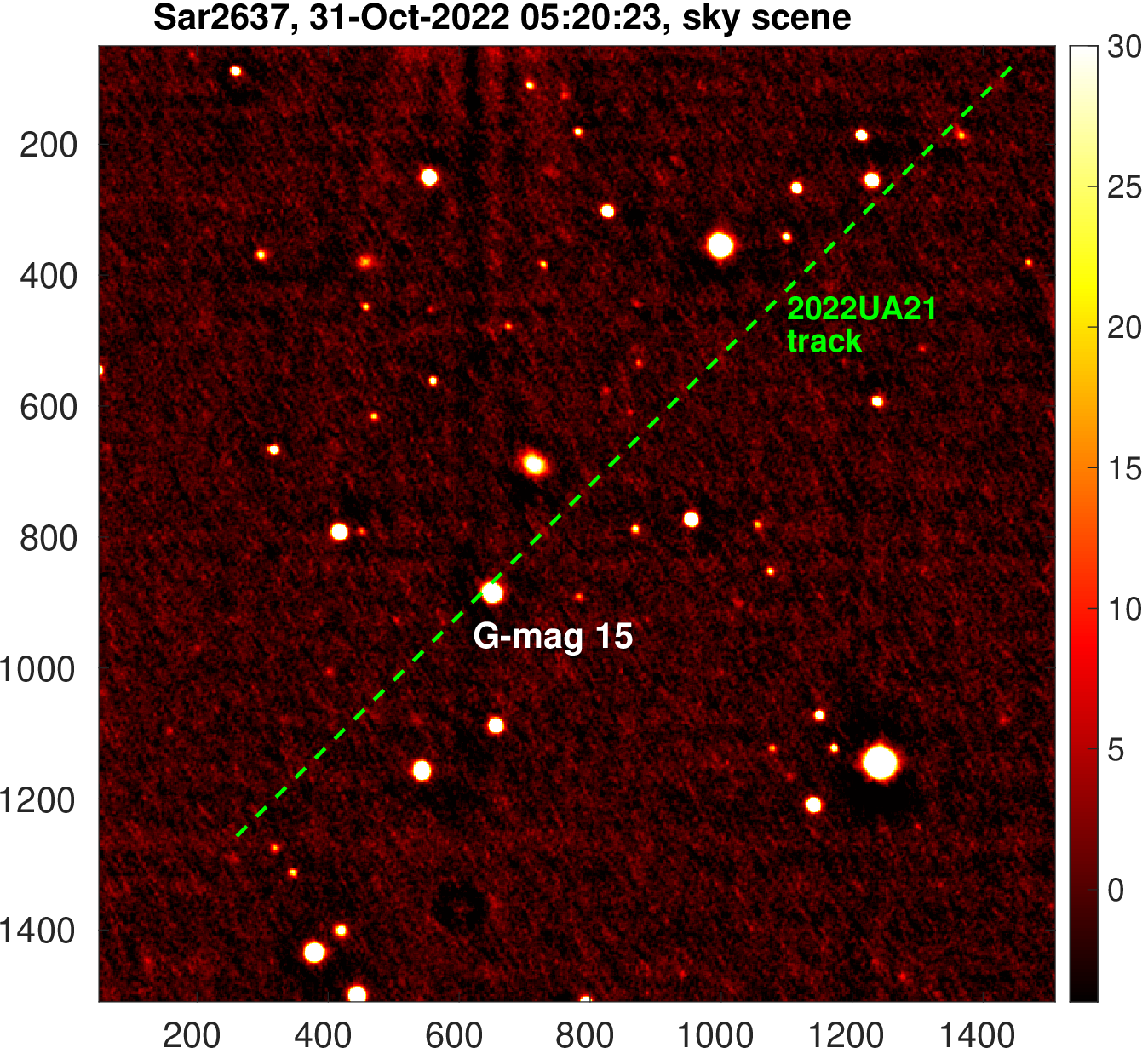}
\epsscale{0.30}
\plotone{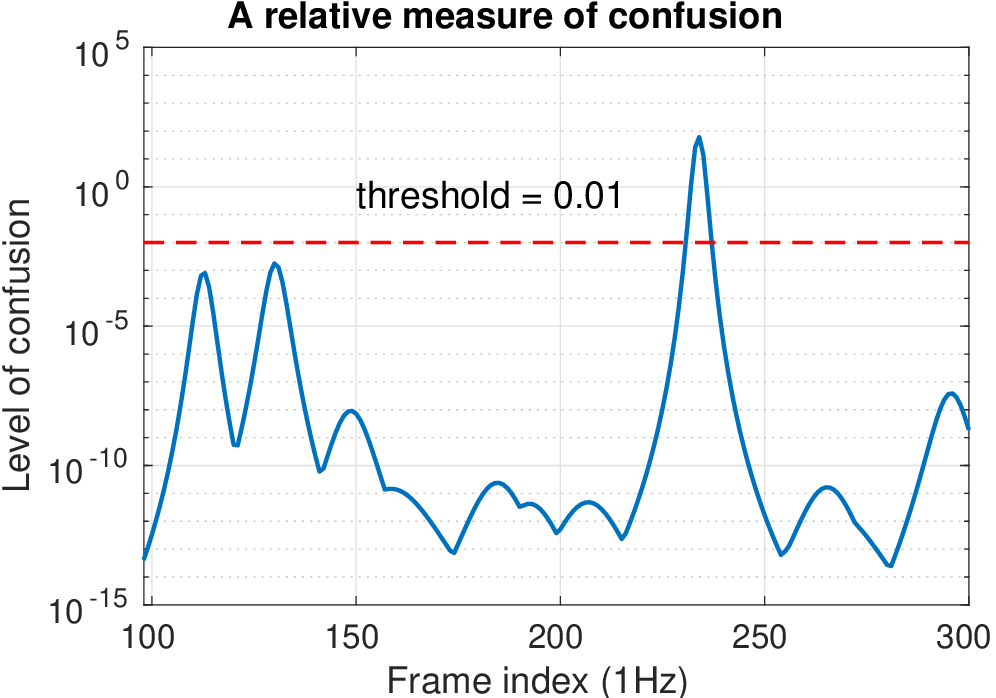}
\epsscale{0.32}
\plotone{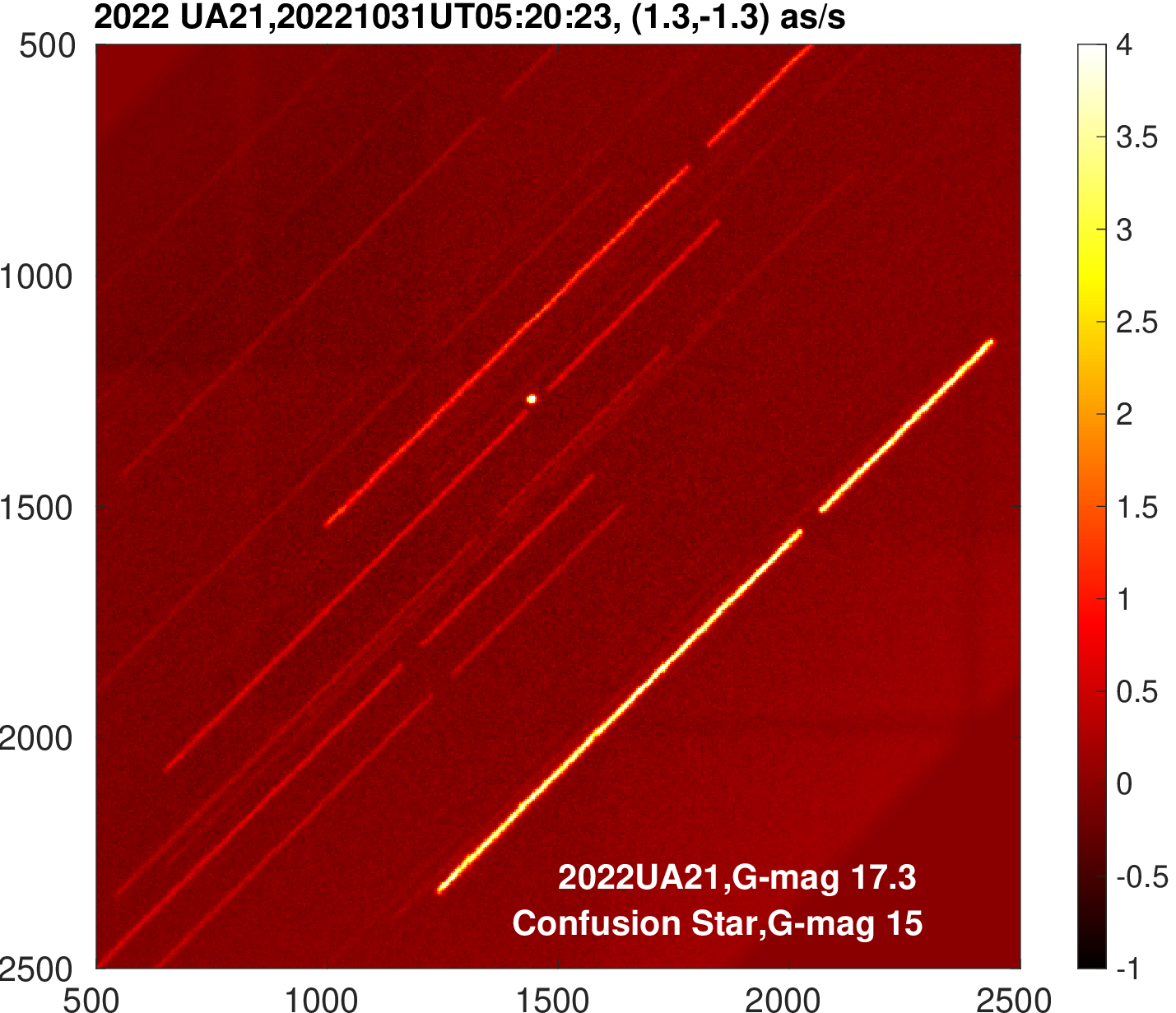}
\caption{Star confusion field (left) and a measure of confusion (right).\label{fig:confusion}}
\end{figure}

\begin{figure}[ht]
\epsscale{0.32}
\plotone{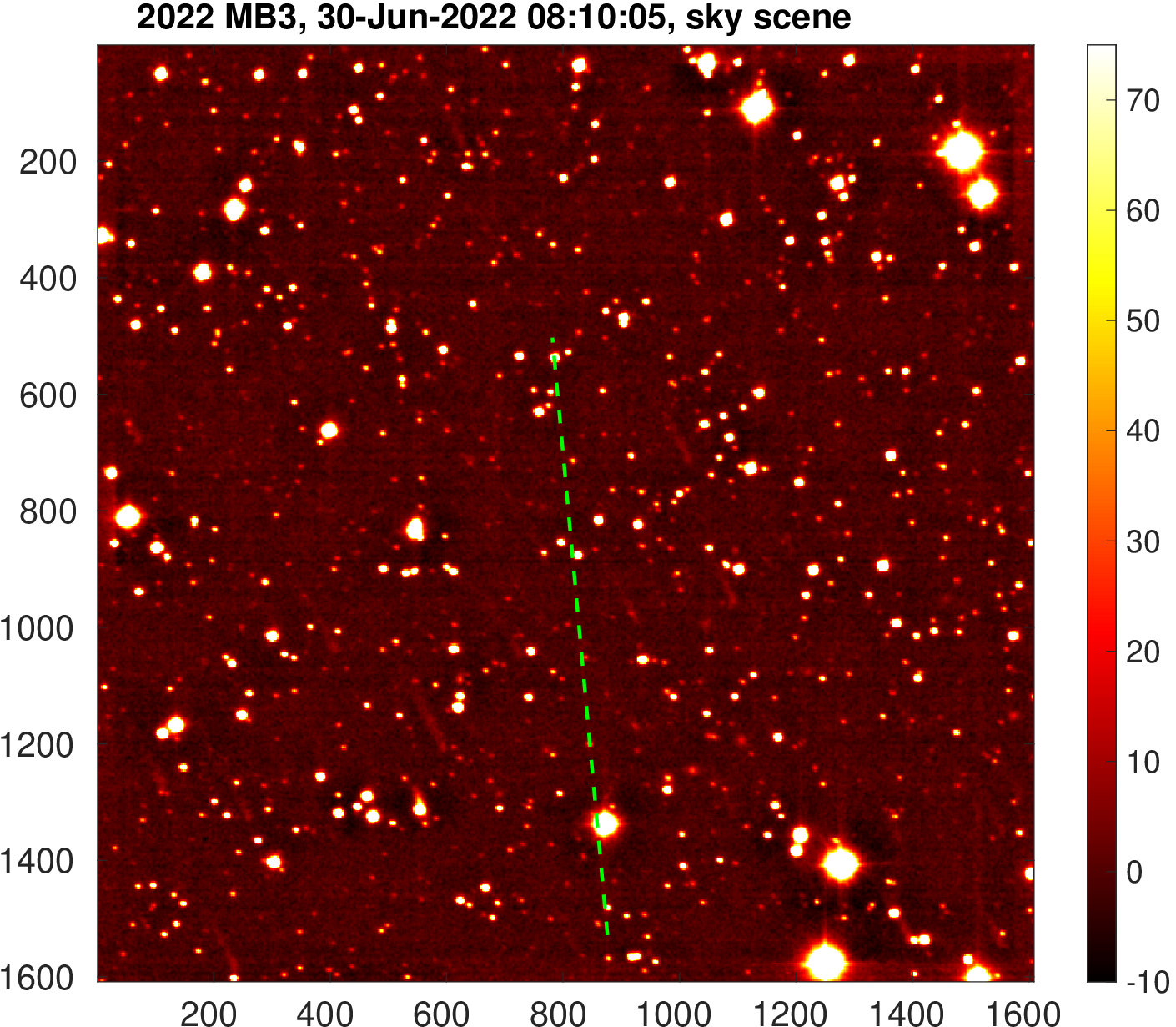}
\epsscale{0.30}
\plotone{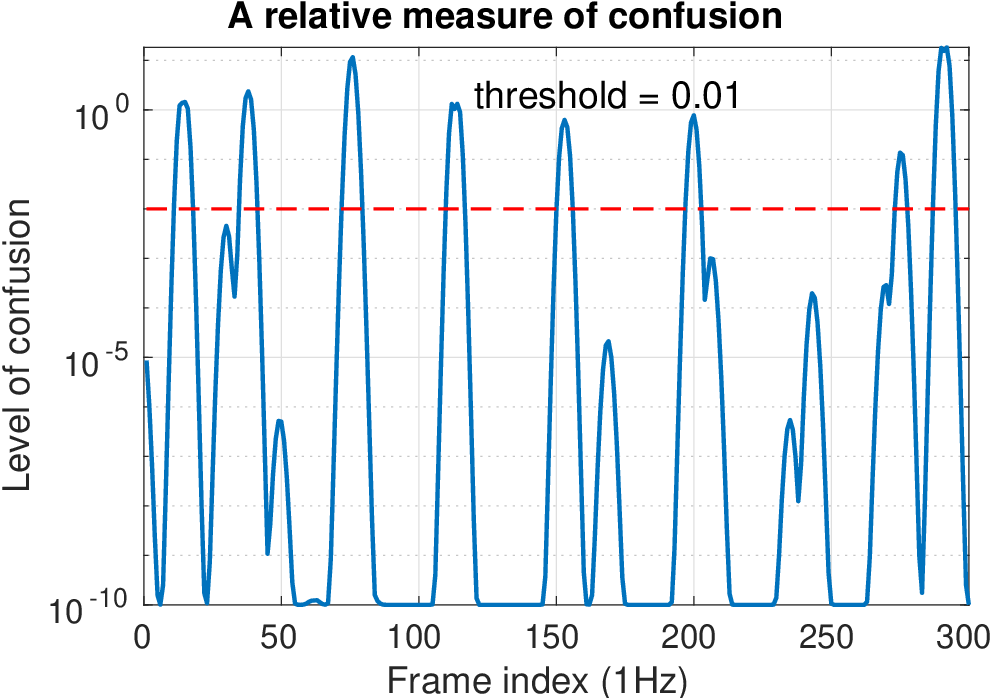}
\epsscale{0.31}
\plotone{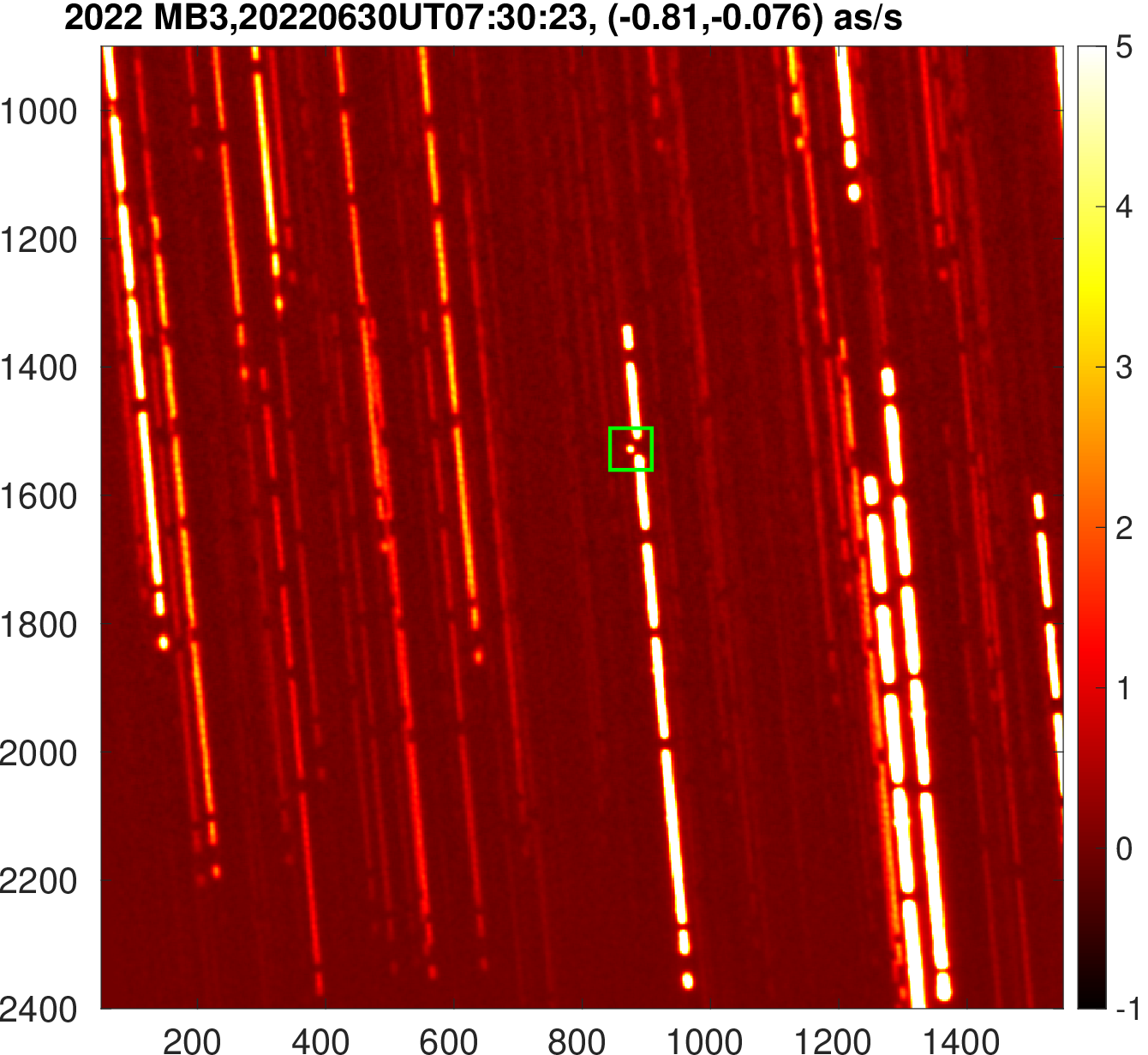}
\caption{Star confusion field (left) and a measure of confusion (right).\label{fig:crowded_confusion}}
\end{figure}

Confusion occurs when a NEO moves very close to a star. If tracking the NEO, the stars streak.
In case of confusion, the streak of a star would overlap with the tracked NEO.
This is an extra burden for operations to avoid confusion, which can be challenging if observing a dense field. 
ST is robust against this confusion because we take short exposures. In post-processing, we can exclude the frames
that have confusion as discussed in subsection~\ref{sec:confusion}.
Fig.~\ref{fig:confusion} shows an example to illustrate our approach. The left plot shows the field with the green dashed line marking the track (from upper right to lower left) 
of NEO 2022UA 21 with an apparent magnitude of 17.3 during our observation.
It encounters a 15-magnitude star, much brighter (about 10 times) than the target. We quantify the confusion by computing the 
intensity gradient of the confusion star at the target relative to the target intensity gradient
(approximately the peak intensity divided by the FWHM) as the measure of the confusion
$\sim |grad(I_{\rm star}(at \; NEO))/ \left ( I_{\rm NEO} (peak)/FWHM \right) $. For this case, the confusion measure is displayed in the mid plot in Fig.~\ref{fig:confusion}.
The right plot shows all the clean frames stacked up tracking the target. Excluding frames with confusion creates gaps in the star streaks so that the target shows in the gap as a compact object without contamination of photons from the stars.
This is particularly useful when the field is crowded as shown in Fig.~\ref{fig:crowded_confusion}.

\subsection{Recovery of Candidates with Highly Uncertain Ephemerides}
Fast-moving NEOs need timely follow-up observations after the initial detection.
NEOs may be discovered without timely follow-up observations due to the unavailability of follow-up facilities, 
poor weather/day-light conditions for observation, or latency in data processing. 
Fast-moving objects tend to develop a large uncertainty quickly due to the propagation of errors. 
Because most of the follow-up facilities do not have large FOVs and the 
vast majority of the facilities rely on tracking the object to avoid trailing loss,
it is quite hard to recover an object that has larger than 1 deg angular uncertainties in the sky position.
Fortunately, our SRO2 system with ST has good capability to recover NEO candidates with relatively large uncertainties
because ST does not require accurate knowledge
of the rate of motion and our 4.47 sqdeg FOV is capable of searching efficiently a large portion of the sky.
To illustrate this capability, Table~\ref{Table3} summarizes four examples of recovery using SRO2,
where we show the large uncertainties of the ephemerides derived from 
the initial discovery in the second column and the last column is the uncertainties after we recover the object.
For example, 2021 TZ13 was discovered on 20211010, based on only 4 observations, the ephemerides are highly uncertain with error 1-5 deg.
Our SRO2 recovered this object, which otherwise would have been lost. Other examples of recovery are for 2022 MD3, 2023 BH5, 2023 BE6.

\begin{table}[H]
\centering
\caption{NEO that are recovered succesfully\label{Table3}}
\begin{center}
\begin{tabular}{|c|c|c|c|c|}
\hline
Asteroid  & Uncertainty at Recovery   & Brightness (mag) & Uncertainty at Next   \\
Designation & (U74) & (mag) & Follow-up Observation \\
 \hline \hline
2021 TZ13 & 1-5 deg & 19.8-20.0 & 2-6 as \\ \hline
2022 MD3 & 5-15 deg & 19.4-20.0 & 2-5 as \\ \hline
2023 BH5&  28-35 deg & 19.5-19.8 &  2-9 as \\ \hline
2023 BE6 & 3-9 deg & 19.8-20.1 & 90-150 as \\ \hline
\end{tabular}
\end{center}
\end{table}

\section{Summary, Discussions, and Future Works}
In summary, ST is effective for observing fast-moving NEOs by avoiding trailing loss to gain detection sensitivity 
and astrometric accuracy. As the field is moving forward quickly,
more small telescopes working with CMOS cameras are used for NEO observations.
These systems are ideal for adopting ST as we have demonstrated with our robotic telescope systems (U68 and U74) at SRO.
Even with the economic COTS hardware, ST is pushing the current state-of-the-art for surveying fast-moving NEOs
to rate faster than (0.5 as/s) and magnitude beyond 20.5.
We recently installed a new system at the Lowell Observatory to have a cluster of 
four 14-inch telescopes (U97), which can be operated in two modes, 
collapsed mode (all the telescopes pointing at the same field of view) and the splayed mode 
where the telescopes point at adjacent fields. 
This is a modern approach for flexibility of using ST to search either deep or a large field.

In the past few years starting the year 2000, the contributions to the total detection from other facilities than the 
major survey facilities like PanSTARRS and CSS have been steadily increasing.
One driving factor is the usage of CMOS with small telescopes and ST. We hope this article will help the community 
use ST and speed up the process of inventorying all the NEOs relevant to planetary defense.

Accurate astrometry provides more accurate future orbital paths for close Earth approaches and more
reliable estimation of probabilities of impacting Earth. 
Another application of accurate ground-based astrometry is in the optical navigation of future spacecraft 
that carry laser communication devices, whose downlink may be used to determine the plane of sky position.

\acknowledgments
The authors would like to thank Heath Rhoades at the Table Mountain Facility of JPL, Tony Grigsby, and Hardy Richardson at the Pomona College for supporting the instrumentation,
and Paul Chodas at JPL for technical advices on using the JPL Horizon System. We thank all the students at the Pomona College who involved in carrying our the NEO observations using the TMF system.
We thank Peter Vere\v{s} at the Minor Planet Center for constantly giving us feedback on our observational data, 
Bill Gray at Project Pluto and Davide Farnocchia  of JPL for helping us improve timing.
This work is supported by NASA's ROSES YORPD program and JPL's internal research fund.
This work has made use of data from the European Space Agency (ESA)
mission {\it Gaia} (\url{https://www.cosmos.esa.int/gaia}), processed by
the {\it Gaia} Data Processing and Analysis Consortium (DPAC,
\url{https://www.cosmos.esa.int/web/gaia/dpac/consortium}). Funding
for the DPAC has been provided by national institutions, in particular
the institutions participating in the {\it Gaia} Multilateral Agreement.
The work described here was carried out at the Jet Propulsion Laboratory, California Institute of Technology,
under a contract with the National Aeronautics and Space Administration.
Copyright 2023. Government sponsorship acknowledged.

\clearpage

\appendix

\section{Detection Signal-to-Noise Ratio for Guassian PSF and Trailing Loss}
In this appendix, we compute the detection SNR for a Gaussian PSF 
\beq
     P_g(x,y) = {1 \over  2 \pi \sigma_g^2} \exp \left (- {x^2 + y^2 \over 2 \sigma_g^2} \right) \,,
\eeq
normalized so that $\sum_{x,y} P_g(x,y) = 1$. We also assume that the PSF is critically sampled.
The total noise in a pixel per frame $\sigma_n$ is 
\beq
    \sigma_n = \sqrt{\sigma_{\rm rn}^2 + \Delta t \left (I_{\rm dark} + I_{\rm bg} \right ) } \,,
    \label{tot_noise}
\eeq
where $\Delta t$ is the exposure time, $I_{\rm dark}$ is the dark current, and $I_{\rm bg}$ is the background flux.
For simplicity and as a good approximation, we consider only uniform background, uniform dark current, and uniform read noise over all the pixels.
For a point source ({\it e.g.}, a star) with flux of $I_{\rm s}$, the counts detected during one exposure 
are $I_{\rm s} P_g(x,y)\Delta t$.
A matched filter with kernel $P_g(x,y)$ gives the highest signal-to-noise ratio (S/N) for detecting the star light.
We calculate the signal as
\beq
    {\rm signal} = I_{\rm s} \Delta t \sum_{x,y} P_g(x,y)^2 \,,
\nonumber
\eeq
and the variance of noise as 
\beq
   Var({\rm noise}) = \sigma_n^2 \sum_{x,y} P_g(x,y)^2 \,,
\nonumber
\eeq
where we have assumed that the noises of pixels are not correlated. Computing 
\beq
\sum_{x,y} P_g(x,y)^2 \approx \int \int dx dy P_g(x,y)^2 = {1 \over 4\pi^2 \sigma_g^4} \int_0^\infty 2 \pi r dr e^{-r^2/\sigma_g^2}={1 \over 4\pi \sigma_g^2} \,,
\eeq
 single frame S/N is
\beq
    {\rm S/N \, (single\;frame)} = {I_{\rm s} \Delta t \over \sigma_n} \sqrt{\sum_{x,y} P_g(x,y)^2} \approx  {I_{\rm s} \Delta t \over \sqrt{4\pi \sigma_g^2} \sigma_n} \,.
\eeq
The full-width-at-half-maximum (FWHM) is commonly used to specify the size of a PSF. FWHM of a Gaussian PSF is given by
\beq
    {\rm FWHM}({\rm Gaussian}) = 2 \sqrt{2 \ln 2} \, \sigma_g \approx 2.355 \sigma_g \,.
\label{fwhm}
\eeq
The S/N for a single frame can be written as
\beq
 {\rm S/N (single\;frame)} \approx  \sqrt{2 \ln 2 \over \pi} {I_s \Delta t \over {\rm FWHM}\; \sigma_n} \approx {I_s \Delta t \over (1.5 \; {\rm FHWM})\, \sigma_n} \,,
\eeq
which means that a Gaussian PSF has the same detection sensitivity as a top-hat square PSF with size $\sim 1.5\times$FWHM.

\begin{figure}[ht]
\epsscale{0.8}
\plotone{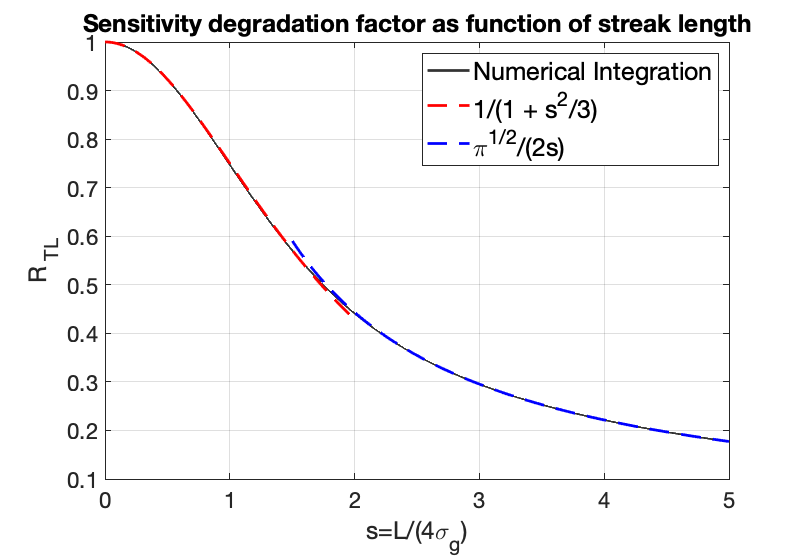}
\caption{Trailing loss sensitivity factor as function of streak length.\label{fig:TL}}
\end{figure}

We now estimate the trailing loss due to motion of the source such as a NEO. 
For a streaked image with streak length $L$, we have the image intensity
\beq
   I(x,y) = {I_{\rm s} \Delta t} {1 \over L} \int_0^L dy' P_g(x, y-y') \,.
\eeq
Ideally, we should use a filter matching $I(x,y)$ to detect a moving object. However, since we do not know the motion in advance, we only use a kernel matching $P_g(x,y)$ for detection.
The convolved signal with a kernel centered at $(x_c, y_c)$ that matches $P_g(x,y)$ is given then
\beq
{\rm signal} (x_c, y_c)= \sum_{x,y} I(x,y) P(x-x_c, y-y_c) \approx {I_{\rm s} \Delta t  \over L} \int_0^L dy' \int dx dy P_g(x-x_c, y-y_c) P_g(x, y-y') \,,
\eeq
Computing 
\beqa
\int dx dy  \!\!\! &P_g &\!\!\!\! (x {-}x_c, y{-}y_c)P_g(x, y{-}y') 
\nonumber\\
\!\!\! &=&\!\!\! {1 \over 4 \pi^2 \sigma_g^4} \int dx dy e^{- \left [ (x - x_c)^2 + (y-y_c)^2 + x^2 + (y-y')^2 \right ]/ (2 \sigma_g^2) }
\nonumber\\
\!\!\!&=&\!\!\!{1 \over 4 \pi^2 \sigma_g^4}  \int dx dy e^{- \left [ (x {-} x_c/2)^2 {+} (y{-}(y_c{+}y')/2)^2 \right ]/ \sigma_g^2} e^{- \left [ x_c^2 {+} (y_c{-}y')^2 \right ]/(4 \sigma_g^2)} 
\nonumber\\
\!\!\!&=&\!\!\!{1 \over 4 \pi  \sigma_g^2}  e^{- \left [ x_c^2 {+} (y_c{-}y')^2 \right ]/(4 \sigma_g^2)} \,,
\eeqa
we get
\beq
  {\rm signal}(x_c, y_c) \approx {I_{\rm s} \Delta t  \over 4 \pi \sigma_g^2  L} \int_0^L dy' e^{- \left [ x_c^2 {+} (y_c{-}y')^2 \right ]/(4 \sigma_g^2)}
\eeq
For the detection of the signal, we only need to consider the maximum of the signal over $(x_c, y_c)$. Therefore, we set $x_c = 0$. In viewing of that the integral over $y'$ reaches
maximum when $y_c = L/2$ by symmetry, {\it i.e.} putting the kernel at the center of the streak. We thus have the maximum signal
\beq
  {\rm signal}\left (x_c{=}0, y_c{=}{L\over 2}\right ) \approx {I_{\rm s} \Delta t  \over 4 \pi \sigma_g^2  L} \int_{-L/2}^{L/2} dy' e^{- y'^2/(4 \sigma_g^2)} = {I_{\rm s} \Delta t  \over 4 \pi \sigma_g^2 } \int_0^1 d\xi e^{- \xi^2(L/(4 \sigma_g))^2}
\eeq
where have changed variable from $y'$ to $\xi \equiv y'/(L/2)$ and used symmetry of the integrand between $+y'$ and $-y'$. 
Therefore,
\beq
  {\rm S/N (streak)} =  {I_{\rm s} \Delta t \over \sqrt{4\pi \sigma_g^2} \sigma_n} \int_0^1 d\xi e^{- \xi^2(L/(4 \sigma_g))^2} ={I_{\rm s} \Delta t \over \sqrt{4\pi \sigma_g^2} \sigma_n} R_{\rm TL}\left (L \over 4 \sigma_g \right)
\eeq
where we have introduced a reduction function $R_{\rm TL}$  due to trailing loss
\beq
R_{\rm TL} (s) \equiv \int_0^1 d \xi e^{-\xi^2 s^2}\,.
\eeq
For small $s$, the following expansion is useful 
\beq
R_{\rm TL} (s) = \int_0^1 d\xi \sum_{n=0}^\infty (-1)^n {\xi^{2n}  s^{2n} \over n!} = \sum_{n=0}^\infty (-1)^n {s^{2n} \over (2n+1) n!} = 1 - {s^2 \over 3} + O(s^4)
\eeq
For large $s$, it is useful to express $R_{\rm TL}(s)$ in terms of the Gaussian error function as
\beq
R_{\rm TL}(s) = {1 \over s} \int_0^s d \xi e^{-\xi^2} = {\sqrt{\pi} \over 2 s} erf(s) = {\sqrt{\pi} \over 2 s} \left [ 1  -  O \left ({e^{-s^2} \over \sqrt{\pi} s} \right ) \right ]
\eeq
Fig.~\ref{fig:TL} plots the $R_{\rm TL}$ as function of $s$.
Looking at the expansion and the curve, we have the following approximation
\beq
  R_{\rm TL} (s) \approx \left \{ 
  \begin{array}{cc}
  &1/(1 + s^2/3)\,, \; {\rm for} \; s < 2 \,, \\ 
  &\sqrt{\pi}/(2 s) \,, \; {\rm for} \; s > 2 \,.
  \end{array}
  \right .
\eeq

\begin{figure}[ht]
\epsscale{0.8}
\plotone{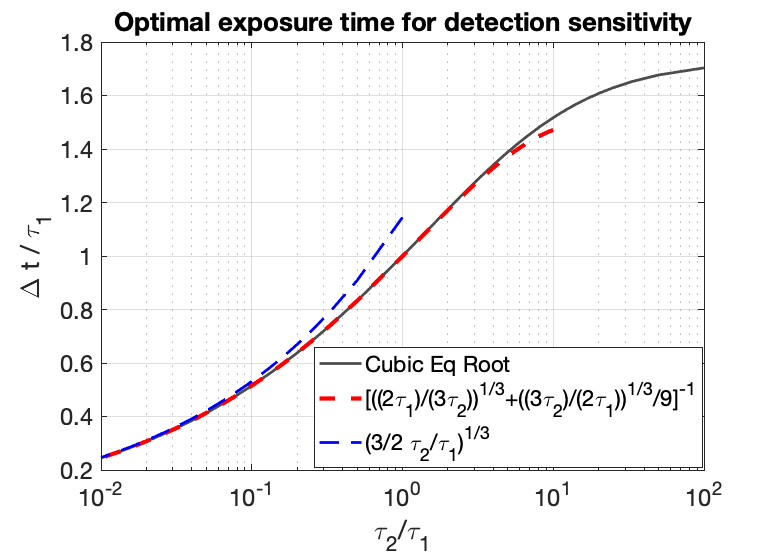}
\caption{Optimal exposure time as function of the ratio of time scales of $\tau_2$ and $\tau_1$. \label{fig:opt_exp}}
\end{figure}

We can now estimate the trailing loss for a streak length equals the FWHM of the PSF, for which $s \approx 0.6$ upon using Eq.~(\ref{fwhm}). 
The loss is roughly $0.6^2/3 \approx 11\%$, so in general keeping the streak length less than the FWHM of the PSF is quite good already.
But, we can further determine the preferred exposure time by maximizing the sensitivity for a fixed integration time $T = N_f  \Delta t$, where $N_f$ is the number of frames.
The total S/N for integrating $N_f$ frames is
\beq
   SNR (\Delta t) \approx \sqrt{N_f} {I_{\rm s} \Delta t \over \sqrt{4\pi \sigma_g^2} \sigma_n} R_{\rm TL}\left (L \over 4 \sigma_g \right)
   \approx  \sqrt {T \over \Delta t}{I_{\rm s} \Delta t \over \sqrt{4\pi \sigma_g^2} \sigma_n}  \left [1 + {1 \over 3} \left ({v \Delta t \over 4 \sigma_g} \right )^2 \right ]^{-1} \,.
\eeq
Inserting Eq.~(\ref{tot_noise}) gives
\beqa
   SNR (\Delta t) &\!\!\! \approx &\!\!\!\! {\sqrt{T \Delta t} I_{\rm s} \over \sqrt{4\pi \sigma_g^2} \sqrt{\sigma_{\rm rn}^2 + \Delta t \left (I_{\rm dark} + I_{\rm bg} \right ) }}
   \left [1 + {1 \over 3} \left ({v \Delta t \over 4 \sigma_g} \right )^2 \right ]^{-1} 
   \nonumber\\
   &\!\!\! = &\!\!\!\! {\sqrt{T } I_{\rm s} \over \sqrt{4\pi \sigma_g^2} \sqrt{\sigma_{\rm rn}^2 / \Delta t+\left (I_{\rm dark} + I_{\rm bg} \right ) }}
   \left [1 +{1 \over 3} \left ({v \Delta t \over 4 \sigma_g} \right )^2 \right ] ^{-1}
\eeqa
It is convenient to introduce time scale, $\tau_1$ for the object to move $4 \sigma_g$, 
\beq
\tau_1 \equiv 4 \sigma_g / v \,.
\eeq
Using Eq.~(\ref{tau2_def}), the dependency of S/N on $\Delta t$ can be expressed as
\beq
    S/N (\Delta t) = {\sqrt{T } I_{\rm s} \over \sqrt{4\pi \sigma_g^2} \sqrt{\left (I_{\rm dark} + I_{\rm bg} \right ) }} \left (1 + {\tau_2 \over \Delta t} \right )^{-1/2}
   \left [1 + {1 \over 3} \left ({ \Delta t \over \tau_1} \right )^2 \right ]^{-1}
\eeq
Taking derivative with respect to $\Delta t$ gives the optimal exposure time $\Delta t$ that maximizes S/N to satisfy the following equation
\beq
 -\left (1 {+ }{\tau_2 \over \Delta t} \right )^{-3/2} \!\! {- \tau_2 \over 2(\Delta t)^2}   \left [1{ +} {1 \over 3} \left ({ \Delta t \over \tau_1} \right )^2 \right ] ^{-1}\!\!\!
 -\left (1 {+} {\tau_2 \over \Delta t} \right )^{-1/2} \left [{2 \over 3 \tau_1} \left ({ \Delta t \over \tau_1} \right ) \right ] \left [1{ +} {1 \over 3} \left ({ \Delta t \over \tau_1} \right )^2 \right ] ^{-2} = 0
\eeq
which can be simplified as 
\beq
  {2 \over 3} \left ( {\tau_1 \over \tau_2} \right ) \left ({ \Delta t \over \tau_1} \right )^3+{1 \over 3} \left ({ \Delta t \over \tau_1} \right )^2 -1 = 0 \,.
\eeq
Using the standard root formula for a cubic equation, we have the solution
\beq
   {\Delta t  \over \tau_1} = \left \{ \left [ {\tau_1 \over 3\tau_2} + \sqrt{\left ( \tau_1 \over 3 \tau_2 \right)^2 - {1 \over 729}} \right ]^{1/3} + 
   \left [ {\tau_1 \over 3\tau_2} - \sqrt{\left ( \tau_1 \over 3 \tau_2 \right)^2 - {1 \over 729}} \right ]^{1/3} \right \}^{-1} \,.
\eeq
In the region where we are dominated by the sky background noise, $\tau_2/\tau_1 \ll 1$, we have the following approximated formula
\beq
   {\Delta t \over \tau_1} \approx \left [ \left (2 \tau_1 \over 3 \tau_2 \right )^{1/3} + {1 \over 9} \left (3 \tau_2 \over 2 \tau_1 \right )^{1/3} \right ]^{-1} \sim \left (3 \tau_2 \over 2 \tau_1 \right )^{1/3} \,, {\rm for} \; \tau_2 \ll \tau_1 \,.
\eeq
Fig.~\ref{fig:opt_exp} shows the optimal exposure time as function of the ratio of the two time scales of $\tau_2$ and $\tau_1$. 
The ratio $\tau_2 / \tau_1$, given by
\beq
{\tau_2 \over \tau_1} = {\sigma_{\rm rn}^2 \over \tau_1 (I_{\rm bg} + I_{\rm dark})} \,,
\eeq
has the physical meaning of the ratio of variances due to read noise and background level integrated over the time of $\tau_1$, 
which is the time for the object to move $4\sigma_g \approx 1.7 \, {\rm FWHM}$.
For applying synthetic tracking, we typically are in the region when $\tau_2 / \tau_1 < 1$ and the streak length is not too large compared with the PSF size.
Since we usually use the FWHM to measure the size of PSF and the streak length instead of using $4 \sigma_g$, we summarize our results for convenient use as the follows.

For short streak length $L < 3\; {\rm FWHM}$, an approximate reduction factor of SNR is given by
\beq
   R_{\rm TL} (L) \approx {1 \over 1 + 1/3 \left ( L \over 4 \sigma_g \right )^2} = {1 \over 1 + \frac{\ln 2}{6} \left ( \frac{L}{FWHM} \right)^2} \approx { 1 \over 1 + 0.115 \left (L \over FWHM \right )^2} \,,
\eeq
For long streak length $ L > 3\; {\rm FWHM}$, sensitivity reduction factor of SNR is given by
\beq
   R_{\rm TL} (L) \approx {2\sqrt{\pi} \sigma_g \over L} = \sqrt{\pi \over 2 \ln 2} \left ( \frac{L}{\rm FWHM} \right )^{-1} \approx 1.5 \left ( \frac{L}{\rm FWHM} \right )^{-1} \,.
\eeq

\section{Differential Chromatic Refraction Correction\label{sec:dcr}}
Because the atmosphere refracts star lights, stars appears as closer to the zenith. The refraction index of the atmosphere depends on the wavelength as $\sim 1/\lambda^2$, thus
the refraction effect of blue stars is larger than that of red stars. For astrometry, we only need to estimate the position of the target relative to reference stars.
If the target and the reference stars all had the same color, the atmospheric refraction effect would be then cancelled up to the field dependent geometric effect, which can be modeled by the field distortion.
However, the target and the reference stars are in general of different stellar types, therefore we need to correct the differential chromatic refraction (DCR) effect. 
This effect can be mitigated by applying a narrow band filter, but this also reduces amount of photons, which may introduce too much photon noises. 
Another way is to limit the stellar type ensure they are close the type of the target, this however will significantly limit the amount of stars thus may lead
to poor astrometric solutions. Fortunately, the DCR effect can be modeled using air refraction index \citep{Stone1996} and the spectra of objects.
For example, \cite{Magnier2020} used a linear color model to correct DCR effects for PanSTARRS1 astrometry calibration.
Here we found, for 10 mas accuracy, this refraction effect can be modeled with a simple quadratic color model:
\beq
\delta (RA, Dec) = \tan(\theta_z) (\sin \phi_z, \cos \phi_z) \left [a C + b C^2 \right ] 
\label{dcr_model2}
\eeq
where $\theta_z$ is the zenith angle, complementary to the elevation angle, $\phi_z$ is the parallactic angle
between the zenith and the celestrial pole from the center of field, and
 $C\equiv B -R $ is the difference of Gaia's blue-pass filter magnitude B and red-pass filter magnitude R
 \citep{Andrae2018}. $a$ and $b$ can be estimated
according a dense field with sufficient stellar spectral diversity. Model~(\ref{dcr_model2}) is the base for Eq.~(\ref{dcr_model}), which
gives the DCR correction accounts for the spectral difference between the reference and target objects in terms of reference star color
 $C_{\rm ref}$ and target color $C_{\rm tar}$.
Note that DCR effect depends on the passband used for observations and this dependency is captured by parameters $a$ and $b$.
To maximize photon usage for sensitivity, we use the full band of the CMOS (``clear filter") by default.
We observed the Freia (76) asteroid using a clear filter and also a Sloan i-band filter. Without and DCR correction,
we have residuals as displayed in the left plot in Fig.~\ref{fig:dcr_off_on} where we can see errors shown as a two-dimensional vector tend to align with the direction pointing to the zenith.
Displaying the component along the zenith direction (altitude) and the direction perpendicular to the zenith direction,
which we call azimuth direction, we found that these errors have a systematic dependency (dominantly linear) with the star colors
$C_{\rm ref} = (B-R)_{\rm ref}$. This supports the model~(\ref{dcr_model2}).
\begin{figure}[ht]
\epsscale{1}
\plottwo{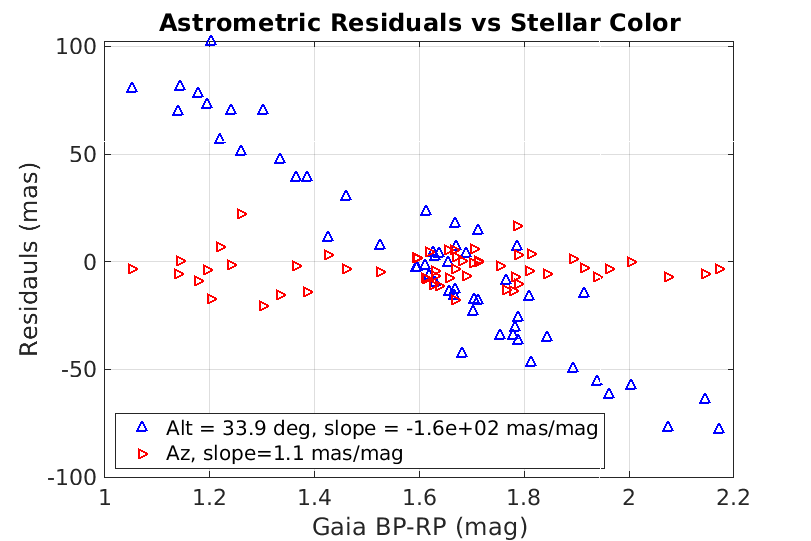}{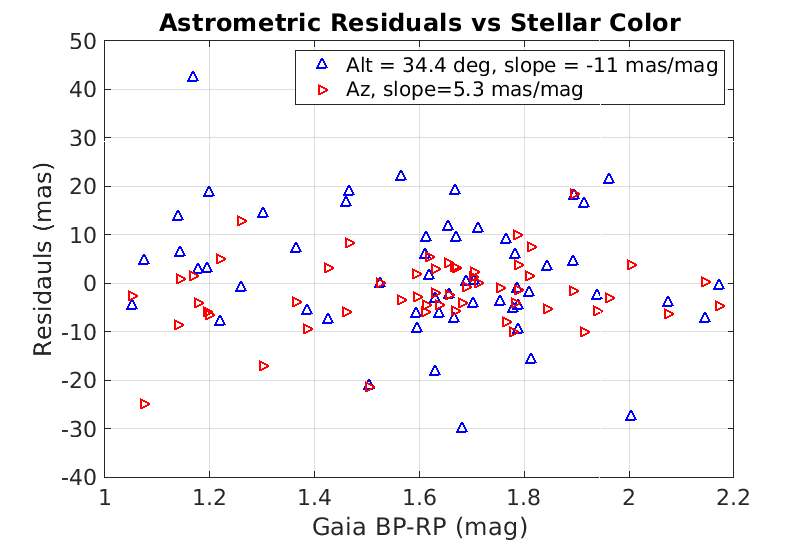}
\caption{Astrometric errors due to DCR effect is approximately a linear function of color in left plot for a clear filter; the dependency is not significant for i-band filter as shown in the right plot.\label{fig:dcr_vs_color}}
\end{figure}
A quadratic fit to this kind of curve allows us to determine coefficients $a$ and $b$ empirically. For example, for our "clear band," $a \approx $-168 mas and $b\approx$ 20 mas. 
In contrast to the clear filter, the residuals shown in Fig.~\ref{fig:astr_sol_i} for the i-band filter is hard to identify and the color dependency is hard to see suggesting that $a$ and $b$ for i-band is smaller than 10 mas.
We also display the astrometric residuals with i-band filter and the DCR effects are much smaller buried in the random noises as shown in the right plot in Fig.~\ref{fig:dcr_vs_color}.

\end{document}